\documentclass[journal]{IEEEtran}
\usepackage{cite}
\usepackage{color}
\usepackage{algorithm}
\usepackage{algpseudocode}
\usepackage{amsfonts}
\usepackage{amsmath}
\usepackage{amsthm}
\usepackage{amssymb}
\usepackage{enumitem}
\usepackage{wrapfig}

\usepackage{graphicx}
\graphicspath{{figure/}}

\usepackage[caption=false,font=footnotesize,labelfont=sf,textfont=sf]{subfig}
\newcommand{\Adv}{\mathcal{A}}
\newcommand{\GC}{\mathcal{GC}}

\newcommand{\Party}{\mathit{P}}
\newcommand{\Client}{\mathit{C}}
\newcommand{\Server}{\mathit{S}}
\newcommand{\Sim}{\mathit{Sim}}

\newtheorem{theorem}{Theorem}
\newtheorem{definition}{Definition}

\begin{document}
\title{Enabling Efficient Privacy-Assured Outlier Detection over Encrypted Incremental Datasets}

\author{Shangqi Lai, Xingliang Yuan, Amin Sakzad, Mahsa Salehi, 
	Joseph K. Liu, and Dongxi Liu
	\thanks{
		S. Lai, X. Yuan, A. Sakzad, M. Salehi, and J.K. Liu are with Faculty of Information Technology, Monash University, Australia, S. Lai and D. Liu are also with Data61, CSIRO, Melbourne/Sydney, Australia. Email Addresses: shangqi.lai,xingliang.yuan,amin.sakzad, mahsa.salehi,joseph.liu@monash.edu, dongxi.liu@data61.csiro.au
	}
	\thanks{Copyright (c) 2019 IEEE. Personal use of this material is permitted. However, permission to use this material for any other purposes must be obtained from the IEEE by sending a request to pubs-permissions@ieee.org.}
}

%\author{First Author\inst{1}\orcidID{0000-1111-2222-3333} \and
%Second Author\inst{2,3}\orcidID{1111-2222-3333-4444} \and
%Third Author\inst{3}\orcidID{2222--3333-4444-5555}}
%\institute{Princeton University, Princeton NJ 08544, USA \and
%Springer Heidelberg, Tiergartenstr. 17, 69121 Heidelberg, Germany
\maketitle              % typeset the header of the contribution
\begin{abstract}
		Outlier detection is widely used in practice to track the anomaly on incremental datasets such as network traffic and system logs.
	However, these datasets often involve sensitive information, and sharing the data to third parties for anomaly detection raises privacy concerns.
	In this paper, we present a privacy-preserving outlier detection protocol (PPOD) for incremental datasets.
	%
	%Our protocol uses the two-server model where the data owner shares their dataset to two untrusted but non-colluding servers to model and track the outliers via the two-party computation technique.
	%
	The protocol decomposes the outlier detection algorithm into several phases and recognises the necessary cryptographic operations in each phase.
	It realises several cryptographic modules via efficient and interchangeable protocols to support the above cryptographic operations and composes them in the overall protocol to enable outlier detection over encrypted datasets.
	To support efficient updates, it integrates the sliding window model to periodically evict the expired data in order to maintain a constant update time.
	We build a prototype of PPOD and systematically evaluates the cryptographic modules and the overall protocols under various parameter settings.
	Our results show that PPOD can handle encrypted incremental datasets with a moderate computation and communication cost.\end{abstract}

%\keywords{Outlier Detection  \and Secure Computation.}

\section{Introduction}
The increasing demands of local and global secrecy and private computations inside a cloud atmosphere reveal essential requirements and commitments to technologies attaining a high level of security. 
In the past few years, the advances in technologies related to the Internet of Things (IoT) has led to a boom in a broad spectrum of areas such as cloud computing, data mining, and information security. 
In particular, cloud service providers are to remove the burden of data management using cost-efficient data mining approaches. 
Hence, it is quite natural for both individuals and organisations to outsource their information data into a cloud server and allow this entity to process the data and run different data mining algorithms on the user's behalf. 
However, storing/processing sensitive data on untrusted cloud servers may raise serious privacy and security concerns for time-series related data in IoT applications.

One of the significant data processing tasks in IoT applications is anomaly detection (outlier detection).\footnote{We use these two terms interchangeably in our paper.} 
Anomaly detection is the process of finding unusual patterns in data, and it has many applications in~\cite{sadik2014research}, intrusion detection~\cite{yuan2016privacy}, and fraud detection~\cite{chandola2009anomaly}. 
In the context of IoT devices, the anomaly detection can be used to remotely detect the malicious behaviours of IoT sensors, which are compromised by attackers~\cite{fu2018risks}. 
The generated data is incremental/temporal (time-series data) and the volume of data to be analysed is effectively large and unbounded \cite{sadik2014research,salehi2016fast}. 
Hence, an anomaly detection algorithm in this setting should be efficient in terms of computational costs and effective in terms of detection accuracy. 
While encryption can be used to address data privacy issues, it prevents the server from mining/processing the encrypted data unconditionally. 
In this paper, we shall propose a new mechanism ``Privacy-Preserving Outlier Detection (PPOD)'' that addresses the problem of mining on encrypted data efficiently and effectively. 
Moreover, in order to make the process of anomaly detection more effective, we aim to consider the temporal relationships in time-series by leveraging the ideas in autoregression forecasting models in the context of uni/multivariate time-series. 
These models can detect deviation based anomalies by considering temporal relationships of measurements in time-series~\cite{gupta2014outlier}. 
The privacy-preserving anomaly detection is a significant area of research and none of the state-of-the-art techniques has addressed it in the presence of temporal data.

Our system architecture comprises of a user (Gateway) and two honest but non-colluding servers in charge of performing secure outlier detection (See Fig.~\ref{fig:overview}).
A PPOD for an incremental dataset $D$ contains four algorithms: 
(1) Data Preprocessing: to generate an encrypted incremental dataset ED and distribute shares of received data points to servers. 
(2) Initialisation: to apply forms of secure multiparty computations to model the outliers of ED. 
This phase outputs the initial list of $k$-distances. 
(3) Query: to run a data mining algorithm on servers and detect anomalies associated with ED in a privacy-assured form. 
(4) Update: to take into account the newly arrived data points, compute their $k$-distances, and decide if they are anomaly or not.
Hence, the specific contributions of this work are as follows:
\begin{itemize}
	\item design a PPOD scheme based on well-known cryptographic protocols/primitives such as additive secret sharing and Yao's garbled circuit and efficient data mining anomaly detectors such as kNN suitable for current IoT cloud services.
	We also prove that our PPOD scheme is secure with a given leakage function in a hybrid model, where parties are given access to the trusted party computing the ideal function of oblivious transfer (OT)~\cite{rabin2005exchange}.	
	\item to handle incremental datasets efficiently, our PPOD incorporates the sliding window model and adapts proper plaintext outlier detection algorithms~\cite{angiulli2007detecting, kontaki2011continuous} for streams for efficient and secure outlier detection.
	\item implement such a construction using computer simulations and analyse its accuracy and efficiency on incremental datasets for different system parameters. 
	Our evaluations on a real-world dataset with $4200$ 16-dimensional data points show that PPOD has a practical performance: it can answer outlier queries within $217$ ms and take $9$ s to update the outlier model after receiving a new data point.
\end{itemize}

\noindent{\bf Organisation.}
The rest of this chapter is structured as follows. We discuss related work in Section~\ref{sec:related}. 
In Section~\ref{sec:pre}, we introduce the background knowledge of distance-based outlier detection algorithms and the needed cryptographic primitives. 
Then, we describe the system overview and its threat model in Section~\ref{sec:overview}.
A detailed construction of cryptographic modules and protocols is presented in Section~\ref{sec:construction}. 
In Section~\ref{sec:security}, we briefly discuss the security of PPOD.
Next, we describe our prototype implementation and evaluation results in Section~\ref{sec:evaluation}. 
We give a conclusion in Section~\ref{sec:conclusion}.

\section{Related Work}\label{sec:related}
	{\noindent\bf Privacy-preserving outlier detection.} The research in privacy-preserving outlier detection has two main streams, i.e., differential privacy-based approaches~\cite{bhaduri2011privacy,bohler2017privacy,erfani2014privacy} and cryptographic-based (secure computation-based) approaches~\cite{alabdulatif2017privacy,li2015privacy,vaidya2004privacy}.
	The differential privacy-based approaches rely on the data perturbation technique to add noise to protect the inputs from the multi-party~\cite{bhaduri2011privacy}.	
	To address the collusion issue in~\cite{bhaduri2011privacy}, Random Multiparty Perturbation (RMP) technique~\cite{erfani2014privacy} is proposed to allow each party to use a unique and different perturbation matrix to randomise their data.
	A recent differential privacy-based work~\cite{bohler2017privacy} leverages a relaxed version of differential privacy to process the data in data streams.
	However, the differential privacy-based approaches lead to an accuracy loss in practice, while our PPOD does not degrade the accuracy comparing to the outlier detection algorithm for unencrypted datasets.
	The secure computation-based approaches are devised via Yao's Garbled Circuit~\cite{vaidya2004privacy}, Homomorphic Encryption~\cite{alabdulatif2017privacy} and the hybrid approach like in this paper~\cite{li2015privacy}.
	Note that the above approaches are designed for the multi-party setting, i.e., each party has its private input, which are not suitable in the application scenario of this paper (outsourced outlier detection).
	
	{\noindent\bf Distance-based Outlier detection for incremental datasets (data streams).} A large number of outlier detection algorithms (e.g.~\cite{angiulli2007detecting,cao2014scalable,kontaki2011continuous}) are proposed to support efficient outlier detection over the incremental datasets (or data streams). 
	However, all the above algorithms only can process the data in an unencrypted form.
	Furthermore, these algorithms involve range queries which has multiple dedicated attacks for its encrypted version~\cite{grubbs2017leakage,kornaropoulosdata} recently.

\section{Preliminaries}\label{sec:pre}
\subsection{Distance-based Outlier Detection}\label{sec:DOD}

\begin{figure}[!t]%{0.4\linewidth}
	\label{fig:outlier}
	\centering
	%\vspace{-1cm}
	\includegraphics[width=0.6\linewidth]{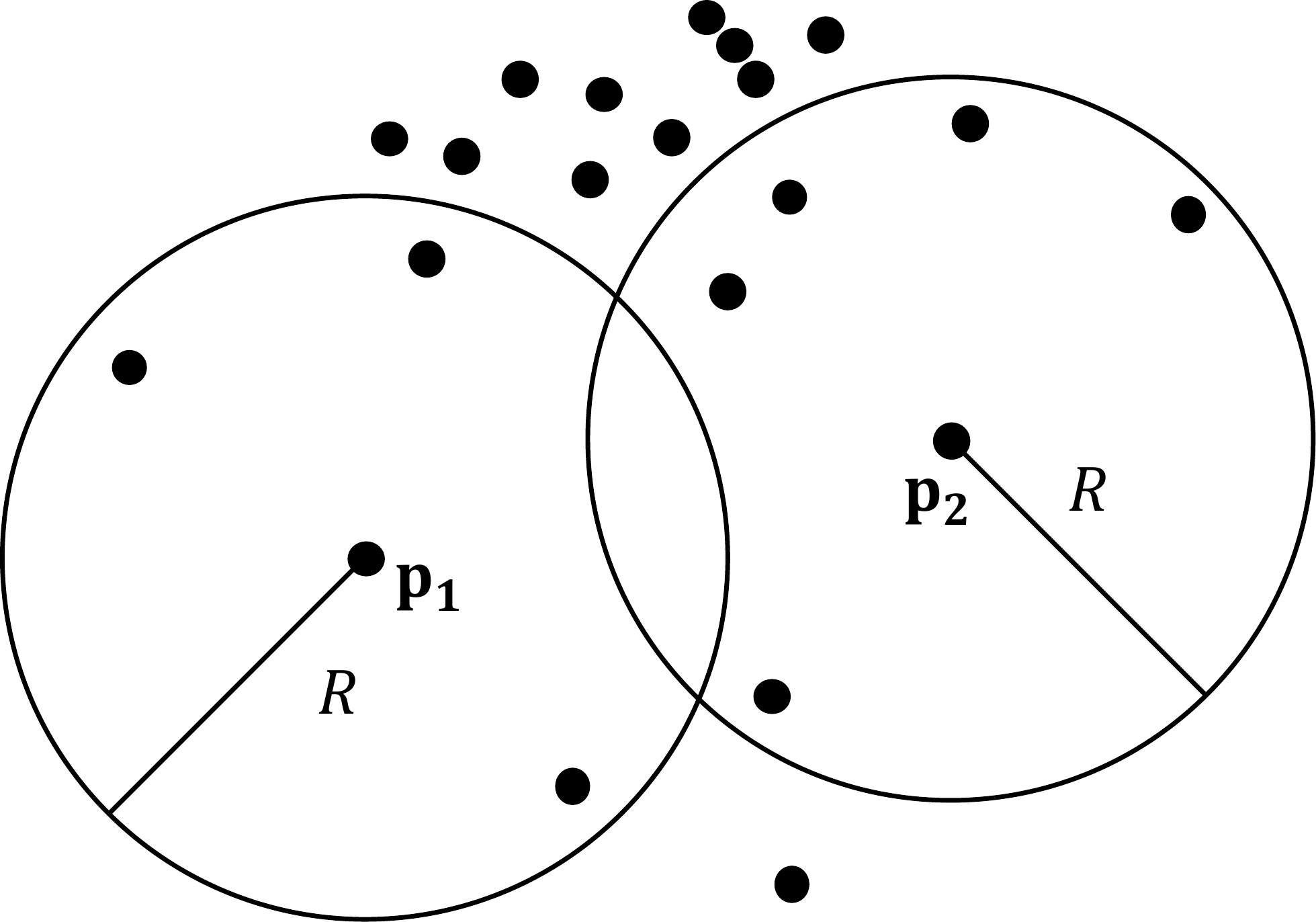}
	\caption{An example of the distance-based outlier}
\end{figure}
We briefly review the formal definitions of distance-based outlier detection. A more detailed introduction can be found in~\cite{tran2016distance}.

Distance-based outlier detection aims to detect an abnormal data point (a.k.a., outliers) via a distance measure between the target point and other points  in a given dataset.
In particular, a neighbour of an $n$-dimensional data point $\mathbf{p}=(p_1, ..., p_n)$ in the distance-based approach is defined as follows.
\begin{definition}[Neighbour]
	Given a distance threshold $R>0$, a data point $\mathbf{q}$ is a neighbour of the target point $\mathbf{p}$ if the distance $\mathsf{d}(\mathbf{p}, \mathbf{q})$ between them is not greater than $R$, where $\mathsf{d}(\cdot, \cdot)$ is a distance measurement function.
\end{definition}
In the distance-based outlier detection approach, normal data points are assumed to have a dense neighbourhood while outliers are far apart from their neighbours (i.e., have a sparse neighbourhood).
Therefore, the distance-based approach utilises the number of neighbours to detect outliers in a dataset.
\begin{definition}[Distance-based Outlier]
	Given a dataset $D$ and a positive integer count threshold $k$, a data point $\mathbf{p}$ is a distance-based {\bf outlier} in $D$ if it has less than $k$ neighbours. Otherwise, it is called an {\bf inlier}.
\end{definition}
Fig.~\ref{fig:outlier} depicts a scenario where the distance threshold $R$ is fixed and $k=5$. According to the above definition, a point $\mathbf{p}$ is an outlier if there are less than $5$ points within the distance $R$ from $\mathbf{p}$ (excluding $\mathbf{p}$ itself). 
Thus, $\mathbf{p_1}$ is an outlier while $\mathbf{p_2}$ is an inliner in the example.

\subsection{Outlier Detection for Incremental Datasets}
The distance-based outlier detection can be exploited to detect outliers in an incremental dataset too, where the dataset is continuously updated with the newly-presented data points.
In this work, we adopt the so-called count-based window model as in the previous works~\cite{angiulli2007detecting, kontaki2011continuous}.
\begin{definition}[Count-based Sliding Window]
	Given a window size $W$ and a slide size $S$. Each window has a starting count $C_{start}$ and an ending count $C_{end}=C_{start}+W$. The window `slides' periodically after receiving a specific number of new data points, causing $C_{start}$ and $C_{end}$ to increase by $S$.
\end{definition}
\noindent In this sliding window model, each data point is associated with a counting number $C_{\mathbf{p}}$. A data point $\mathbf{p}$ is active if its counting number satisfies the following $C_{start}< C_{\mathbf{p}}\leq C_{end}$ (i.e., $W$ active points).

To detect outliers over incremental datasets, a naive solution is to re-compute the neighbours for all active points when the window slides, which can be computationally expensive.
Thus, recent studies devised incremental algorithms: during the update, only the data points which have at least one added/expired neighbour will be updated.
In particular, those algorithms~\cite{angiulli2007detecting, kontaki2011continuous} involve two steps:
\begin{itemize}[leftmargin=*]
	\setlength{\itemsep}{0pt}
	\setlength{\parsep}{0pt}
	\setlength{\parskip}{0pt}
	\item {\bf Expired slide processing:} Data points in the expired slide are removed from the outlier set $\mathbf{O}$ and data point set $\mathbf{P}$.
	However, the expired point can still resident in the neighbour list of active points~\cite{angiulli2007detecting}. 
		
	\item {\bf New slide processing}: For each new data point $\mathbf{p'}$, the algorithm computes its neighbourhood information to determine whether $\mathbf{p}'$ is an outlier or not ($\#$ of neighbours of $\mathbf{p'}\geq k$). 
	Then, for each neighbour point $\mathbf{p}$ of $\mathbf{p'}$,  the neighbour information will be updated regarding the newly-added distance (if $\mathsf{d}(\mathbf{p},\mathbf{p'})\leq R$, then $\mathbf{p}$  has one new neighbour).
	Finally, the algorithm rechecks $\mathbf{p}$ to decide its outlier status according to the new neighbourhood information of $\mathbf{p}$.
\end{itemize}

Note that PPOD follows the above two steps to update the outlier model securely. Thus, it will not incur accuracy loss compared to the algorithms for plaintext data.

\subsection{Secure Computation}\label{subsec:secureComp}
We briefly review the secure computation technologies used in this paper. 
Furthermore, we introduce the secure conversion method, which helps to mix efficient, secure protocols for different computations (e.g., addition, multiplication and sorting) together to support complex computations that involved in our secure outlier detection protocol efficiently.
Readers can find a more detailed introduction in~\cite{pullonen2012design,demmler2015aby}.

{\bf Additive sharing and multiplication triplets.} To additively share ($\mathsf{Shr^A}(\cdot)$) an $\ell$-bit integer $a$ between two parties $\Party_0$ and $\Party_1$, the client generates $a_0\in \mathbb{Z}_{2^l}$ uniformly at random and computes $a_1=a-a_0 \mod 2^l$. 
The first party's share is denoted by $\langle a\rangle_0^A=a_0$ and the second party's is $\langle a\rangle_1^A=a_1$, the modulo operation is omitted in the description later. 
To reconstruct ($\mathsf{Rec^A}(\cdot,\cdot)$) a shared value $\langle a\rangle^A$, each party sends its share to the client who computes $\langle a\rangle_0^A+\langle a\rangle_1^A$.
Given two shared values $\langle a\rangle^A$ and $\langle b\rangle^A$, Addition ($\mathsf{Add^A}(\cdot,\cdot)$) is easily performed non-interactively. 
In detail, $\Party_i$ locally computes $\langle c\rangle_i^A=\langle a\rangle_i^A+\langle b\rangle_i^A \mod 2^l, i\in\{0,1\}$, which also can be denoted by $\langle c\rangle^A=\langle a\rangle^A + \langle b\rangle^A$. 

\noindent To multiply ($\mathsf{Mul^A}(\cdot,\cdot)$) two shared values $\langle a\rangle^A$ and $\langle b\rangle^A$, we leverage Beaver's multiplication triplets technique~\cite{beaver1991efficient}. 
Assuming that the two parties have already precomputed and shared $\langle x\rangle^A$, $\langle y\rangle^A$ and $\langle z\rangle^A$, where $x,y$ are uniformly random values in $\mathbb{Z}_{2^l}$, and $z=x\cdot y\mod 2^l$. 
Then, $\Party_i$ computes $\langle e\rangle_i^A=\langle a\rangle_i^A-\langle x\rangle_i^A$ and $\langle f\rangle_i^A=\langle b\rangle_i^A-\langle y\rangle_i^A$. 
Both parties run $\mathsf{Rec^A}(\langle e\rangle_0^A, \langle e\rangle_1^A)$ and $\mathsf{Rec^A}(\langle f\rangle_0^A, \langle f\rangle_1^A)$ to get $e$ and $f$, and $\Party_i$ lets $\langle c\rangle_i^A=i\cdot e\cdot f+f\cdot\langle x\rangle_i^A+e\cdot \langle y\rangle_i^A+\langle z\rangle_i^A, i\in\{0,1\}$.

\noindent{\bf Garbled circuit and Yao's sharing.} Yao's Garbled Circuit (GC) is first introduced in~\cite{yao1982protocols}, and its security model has been formalised in~\cite{bellare2012foundations}. 
GC is a generic tool to support secure two-party computation. 
The protocol is run between a ``garbler'' with a private input $x$ and an ``evaluator'' with its private input $y$. The above two parties wish to securely evaluate a function $f(x,y)$. At the end of the protocol, both parties learn the value of $z=f(x, y)$, but no party learns more than what is revealed from this output value. 
%
%In details, the garbler runs a garbling algorithm $\GC$ to generate a garbled circuit $F$ and a decoding table $dec$ for function $f$. The garbler also encodes its input $x$ to $\hat{x}$ and sends it to the evaluator. The evaluator runs an oblivious transfer (OT)~\cite{asharov2013more} protocol with the garbler to acquire its encoded input $\hat{y}$. Finally, the evaluator can compute $\hat{z}$ from $F, \hat{x}, \hat{y}$, decode it with $dec$, and share the result $z$ with the garbler. 
%
%The security proof against a semi-honest adversary under two-party setting is given in~\cite{lindell2009proof}.

In the rest of this paper and without loss of generality, we assume that $\Party_0$ is the garbler and $\Party_1$ is the evaluator.
GC can also be considered as a protocol which takes as inputs the Yao's shares and produces the Yao's shares of outputs. 
In particular, the Yao's shares of 1-bit value $a\in\{0,1\}$ is denoted as $\langle a\rangle_0^Y=\{K_0, K_1\}$ and $\langle a\rangle_1^Y=K_a$, where $K_0, K_1$ are the labels representing $0$ and $1$, respectively. 
The garbler runs a garbling algorithm $\GC$ to generate the garbled circuit and its encoded inputs in the form of Yao's shares.
Then, the garbler sends the Yao's shares corresponding to its input to the evaluator.
Meanwhile, the evaluator runs an oblivious transfer (OT)~\cite{asharov2013more} protocol with the garbler to acquire the Yao's shares corresponding to its input.
Then, the evaluator uses the received shares to evaluate the generated circuit and gets the output shares (other labels).

\noindent{\bf Conversion.} Secure computations based on the above two schemes can be combined by converting one representation of intermediate values to the other~\cite{demmler2015aby}. 
Additive shares can be switched to Yao's shares ($\mathsf{A2Y(\cdot)}$) efficiently. 
To be more precise, two parties share their additive shares $a_0=\langle a\rangle_0^A$, $a_1=\langle a\rangle_1^A$ in a bitwise fashion via Yao's sharing. 
The evaluator then receives $\langle a_0\rangle^Y$ and $\langle a_1\rangle^Y$ and evaluates the circuit $\langle a_0\rangle^Y+\langle a_1\rangle^Y$ to get the label of $a$.
Similarly, Yao's shares of $a$ can be converted to additive shares using a subtraction circuit ($\mathsf{Y2A(\cdot)}$). 
In specific, the garbler chooses a random value $a_0\in \mathbb{Z}_{2^l}$ as $\langle a\rangle_0^A$ and gives the Yao's share of $a_0$ to the evaluator, who evaluates the subtraction circuit $\langle a_1\rangle^Y=\langle a\rangle^Y-\langle a_0\rangle^Y$. 
The evaluator can recover $a_1$ locally and set it as $\langle a\rangle_1^A$.

\section{System Overview}\label{sec:overview}
\subsection{System Architecture}
\begin{figure}[!t]
	\centering
	\includegraphics[width=\linewidth]{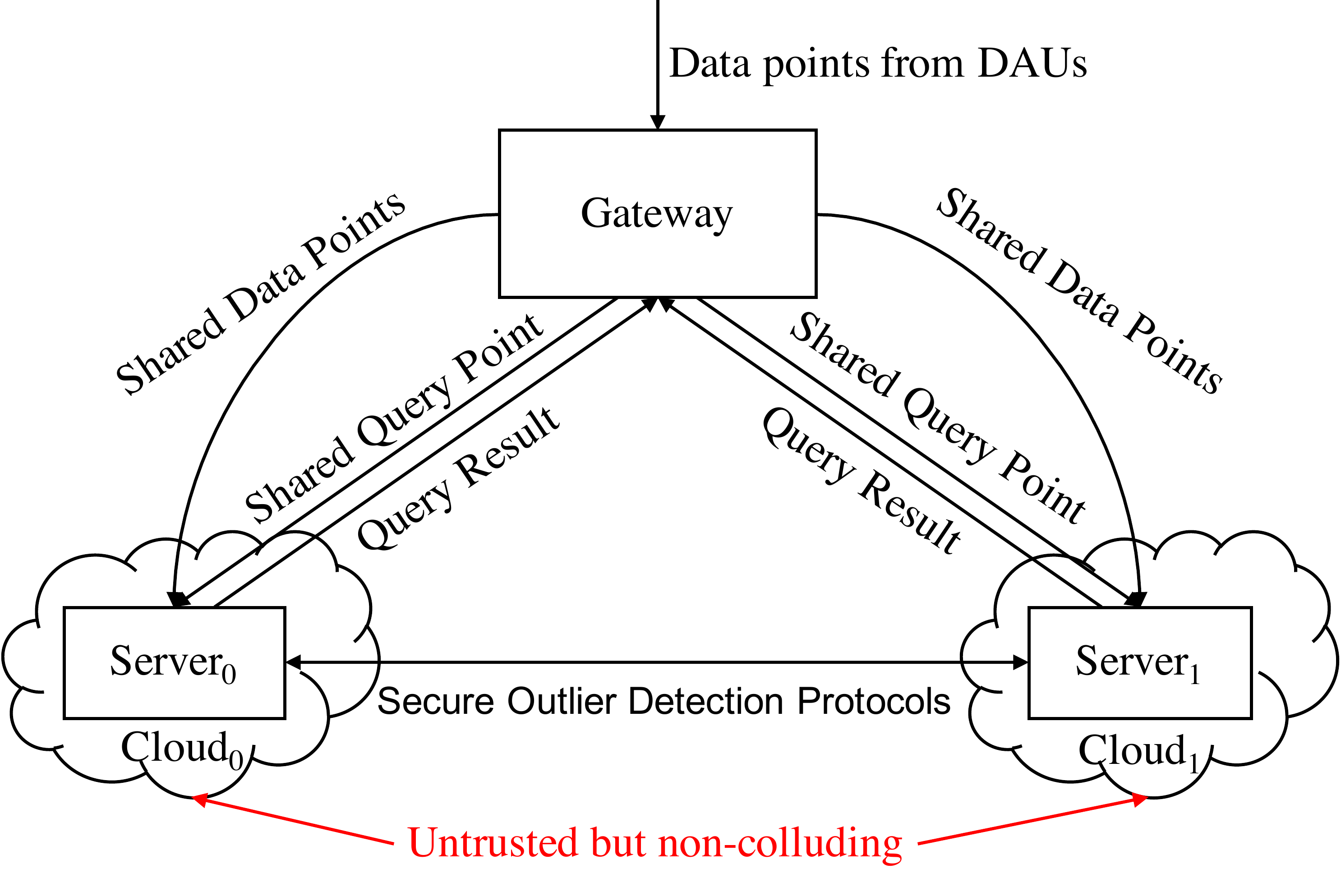}
	\caption{System Overview.}
	\label{fig:overview}
	%\vspace{-10pt}
\end{figure}
Fig.~\ref{fig:overview} shows the system architecture of the PPOD system. 
There are two entities in the PPOD system: the private gateway connected with data acquisition units (DAUs) and the server with the outlier detection service in an untrusted cloud.
Note that this setting reflects the system model of many industrial corporations such as AgentVi~\cite{Agent2018innoVi} and HoneyWell~\cite{Azure2016HoneyWell}, who provide data collection facilities and anomaly detection services while outsourcing the computation part of the service to the cloud service provider. 
In addition, popular cloud providers start to offer the incremental anomaly detection services in their dedicated data mining platform, e.g., Amazon Kinesis~\cite{Amazon2018Kinesis} and Azure Machine Learning Studio~\cite{Microsoft2018Anomaly}.
Our PPOD system aims to protect the confidentiality of the outsourced data in such a trend of using the data analysis cloud platform.

\noindent Our system flow involves four phases: (1) \emph{Data Preprocessing}: For each new data point from DAUs, the gateway preprocesses this point to meet the input requirement of the additive sharing scheme and shares it between two untrusted but non-colluding cloud servers $\Server_0$ and $\Server_1$. 
(2) \emph{Initialisation}: During this phase, the servers execute secure computation protocols to compute $k$-nearest neighbours of each point and to determine the outlier list based on the $k$-distance (i.e., the distance between the data point and its $k$-th nearest neighbour).
Additionally, each server stores the computed $k$-nearest neighbours list and $k$-distances as a reference for the update phase.
(3) \emph{Query}: The user of the PPOD system can submit a query point to the gateway to check whether the point is an outlier or not regarding the current outlier model. 
The query point is also preprocessed and shared by the gateway, and later the server leverages the share to measure the distance between outliers and the query point.
The query point is an outlier in the current model if the computed distance is not greater than an outlier threshold, which is set by the system user.
(4) \emph{Update}: For each new data point, the server follows the same procedure as in the initialisation phase to find the $k$-nearest neighbours of new point and to find the new outlier based on the distance metric ($\mathsf{d}(\cdot,\cdot), R$ and $k$).
Moreover, the server also updates the $k$-nearest neighbours' information of the new-coming data points. In this stage, the server combines the pre-computed information and new distance information to update the $k$-nearest neighbours list for these affected points.
At last, the server refers to the new $k$-nearest neighbours information to decide the status (i.e., outlier or inlier) of these points.

Our system considers a server-aid computation scenario where the internal gateway distributes the computation tasks to two untrusted but non-colluding cloud servers. 
Such a two-server approach has been formalised~\cite{kamara2011outsourcing} and widely utilised in the literature~\cite{mohassel2017secureml,nikolaenko2013privacy,nikolaenko2013ridge,lai2019graph} to protect the data confidentiality in the outsourcing computation context.

\subsection{Threat Model}
In this work, we assume that the gateway and the attached DAUs are maintained by a data analytics service provider, which is a trusted party.
Meanwhile, we consider that the two servers belong to two different semi-honest but non-colluding parties (e.g., two cloud providers). 
They will follow our protocol honestly, but they are interested in learning the underlying private information, which, in our case, are the coordinates of data points.
In the rest of the paper, we use $\Adv_i$ to denote the adversary who compromises $\Server_i$. 
In our security model, we require that the $\Adv_i$ is capable of seeing the protocol messages in $\Server_i$ and tries to infer the user's private information. However, $\Adv_i$ should not learn any information about its counter-party's data beyond the protocol output.
This model aims to protect the confidentiality of data points when the data analytics service providers outsource the computation task to the public cloud.

\section{PPOD Protocol Construction}~\label{sec:construction}
We now explain the construction of PPOD in more details. The notations we used for the algorithms are summarised in Table~\ref{tlb:notation}.

\begin{table}[!t]
	\centering
	\caption{Notations for the outlier detection algorithms.}
	\label{tlb:notation}
	\begin{tabular}{|c|c|}
		\hline
		Notation & Meaning \\ 
		\hline
		$\mathbf{p}$ & A data point with n-dimensional coordinates $(p_1, ..., p_n)$ \\
		\hline
		$\mathsf{d}(\mathbf{p}, \mathbf{q})$ & The distance between data point $\mathbf{p}$ and $\mathbf{q}$ \\
		\hline
		$\mathbf{p}.id$ & The identity of $\mathbf{p}$ \\
		\hline
		$\mathbf{p}.D$ & \begin{tabular}[x]{@{}c@{}}The unordered list of the $k$-nearest neighbours of $\mathbf{p}$\\ in the form of $\{\mathbf{q_i}.id, \mathsf{d}(\mathbf{p}, \mathbf{q_i})\}_{i=1}^k$
 		\end{tabular} \\
		\hline
		$\mathbf{p}.D^k$ & The distance between $\mathbf{p}$ and its $k$-th nearest neighbour \\
		\hline
		$\langle\mathbf{p}\rangle$ & \begin{tabular}[x]{@{}c@{}}$\mathbf{p}$'s secret shares (The coordinates, $D$, $D^k$ and $C_\mathbf{p}$ are\\ stored as secret shares)\end{tabular}\\
		\hline
		$\mathbf{P}$ & A data point set \\
		\hline
		$\langle\mathbf{P}\rangle$ & The data point set keeps the secret shares of data points\\
		\hline
	\end{tabular}
\end{table}

\subsection{Cryptographic Modules}
In order to explain the design clearly, we break the protocol into common used cryptographic modules implemented by the cryptographic primitives (see Section~\ref{sec:pre} for details).
In this section, we discuss the design and implementation of these cryptographic modules. 

\subsubsection{Distance measurement}\label{sec:distance}
In this work, we leverage the squared Euclidean distance to measure the distance between two data points. 
Note that such a distance metric is commonly used in outlier detection algorithms~\cite{chan2005modeling,ramaswamy2000efficient} as it requires less computations (i.e., only addition and multiplication).
However, directly computing $\Sigma_{1\leq i\leq n}(p_i-q_i)^2$ is not applicable in our system due to the non-negative input restriction of additive sharing scheme, i.e., if there exists an $1\leq i \leq n$ such that $p_i-q_i<0$, the square operation will produce the additive shares of an undesired result $(2^l-(p_i-q_i))^2$.
A naive solution for this issue is to make a comparison and swap before computing $p_i-q_i$ to ensure that $p_i-q_i\geq0$, yet it requires additional steps to convert the additive shares to Yao's shares (for comparison) and convert it back (for computation), which leads to extra computation and communication cost.
Therefore, the distance measurement function $\mathsf{d}(\cdot, \cdot)$ in our system is defined as $\mathsf{d}(\mathbf{p}, \mathbf{q})\triangleq \Sigma_{1\leq i\leq n}((p_i)^2+(q_i)^2-2\cdot p_i\cdot q_i)$, which can avoid all negative results as well as the expensive comparison and swap.
Noted that this metric also works when data points are secretly shared. In particular, the two servers can run arithmetic operations to compute their shares as 
$$\langle \mathsf{d}(\langle\mathbf{p}\rangle, \langle\mathbf{q}\rangle)\rangle^A=\Sigma_{1\leq i\leq n}((\langle p_i\rangle^A)^2+(\langle q_i \rangle^A)^2-2\cdot \langle p_i\rangle^A\cdot \langle q_i\rangle^A),$$
independently.
\begin{algorithm}[!t]
	\centering
	\caption{$k$-nearest neighbours}
	\label{alg:knn}
	\begin{algorithmic}[1]
		\Require Shared point set $\langle\mathbf{Q}\rangle$, Shared point $\langle\mathbf{p}\rangle$, Parameter $k$
		\Ensure Yao's shares of the unordered kNN list $L$
		\Function{kNN}{$\langle\mathbf{Q}\rangle$, $\langle\mathbf{p}\rangle$, $k$}
		\State $S\leftarrow \{\}$
		\For{each $\langle\mathbf{q}\rangle\in \langle\mathbf{Q}\rangle$}
		\State $\langle d\rangle^A = \langle \mathsf{d}(\langle\mathbf{p}\rangle, \langle\mathbf{q}\rangle)\rangle^A$
		\State put $\{\langle\mathbf{q}\rangle.id, \langle d\rangle^A\}$ into $S$
		\EndFor
		\State \Return $\textsc{SortShuffle}(S, k)$
		\EndFunction
	\end{algorithmic}
\end{algorithm}
	
\begin{algorithm}[!t]
	\centering
	\caption{$k$-distance}
	\label{alg:kdist}
	\begin{algorithmic}[1]
		\Require Yao's shares of the kNN list $L$
		\Ensure Yao's shares of the maximum value in $L$
		\Function{kDist}{$L$}		
		\State \Return $\textsc{Max}(L)$
		\EndFunction
	\end{algorithmic}
\end{algorithm}
	
\subsubsection{$k$-nearest neighbours (kNN) and $k$-distance} \label{sec:knn}
To detect outliers in a given set of data points, PPOD employs the distance metric in Section~\ref{sec:distance} to compute the share of distances and utilises these shares to compute the $k$-nearest neighbours (kNN) and $k$-distance of each data point.
Then, it compares the $k$-distance with the parameter $R$; if the $k$-distance is greater than $R$, the data point is an outlier in the current model.
Note that this approach can detect the outliers defined in Section~\ref{sec:DOD}: the $k$-distance of a data point is greater than $R$ is equivalent to the point has less than $k$ neighbours within the given range $R$, and it is an outlier.

The simplest way to securely compute the kNN list and $k$-distance is to retrieve those information from a sorted point list after evaluating a sorting circuit over the share of distances. 
However, the above solution still has two security issues.
First, sorting reveals the order of data points, while some recent works~\cite{grubbs2017leakage,kornaropoulosdata} demonstrated that it is possible to precisely reconstruct the underling values (i.e., distances) if an adversary knows the rank and some auxiliary information.
Furthermore, different kNN lists may include the same data point, which means that the adversary can compare the identities/distance shares in different kNN lists to learn extra information about common neighbours for different data points.
Thus, to protect the privacy of data points, the procedures for kNN and $k$-distance evaluation should not reveal the order as well as the identities/distance shares. 

Algorithm~\ref{alg:knn} and \ref{alg:kdist} outline the overall process of computing kNN and $k$-distance for a point $\mathbf{p}$. 
These algorithms employ two cryptographic sub-modules to implement the secure sorting (\textsc{SortShuffle}) and comparison (\textsc{Max}).
Besides, we provide two other cryptographic modules to preprocess (\textsc{Randomise}) and post-process (\textsc{Derandomise}) the kNN list and $k$-distance to hide the repeat patterns. Fig.~3 summarizes these cryptographic sub-modules.
\begin{figure*}[!t]
	\centering
	\subfloat[SortShuffle]{
		\label{fig:sortshuffle}
		\includegraphics[width=0.5\linewidth]{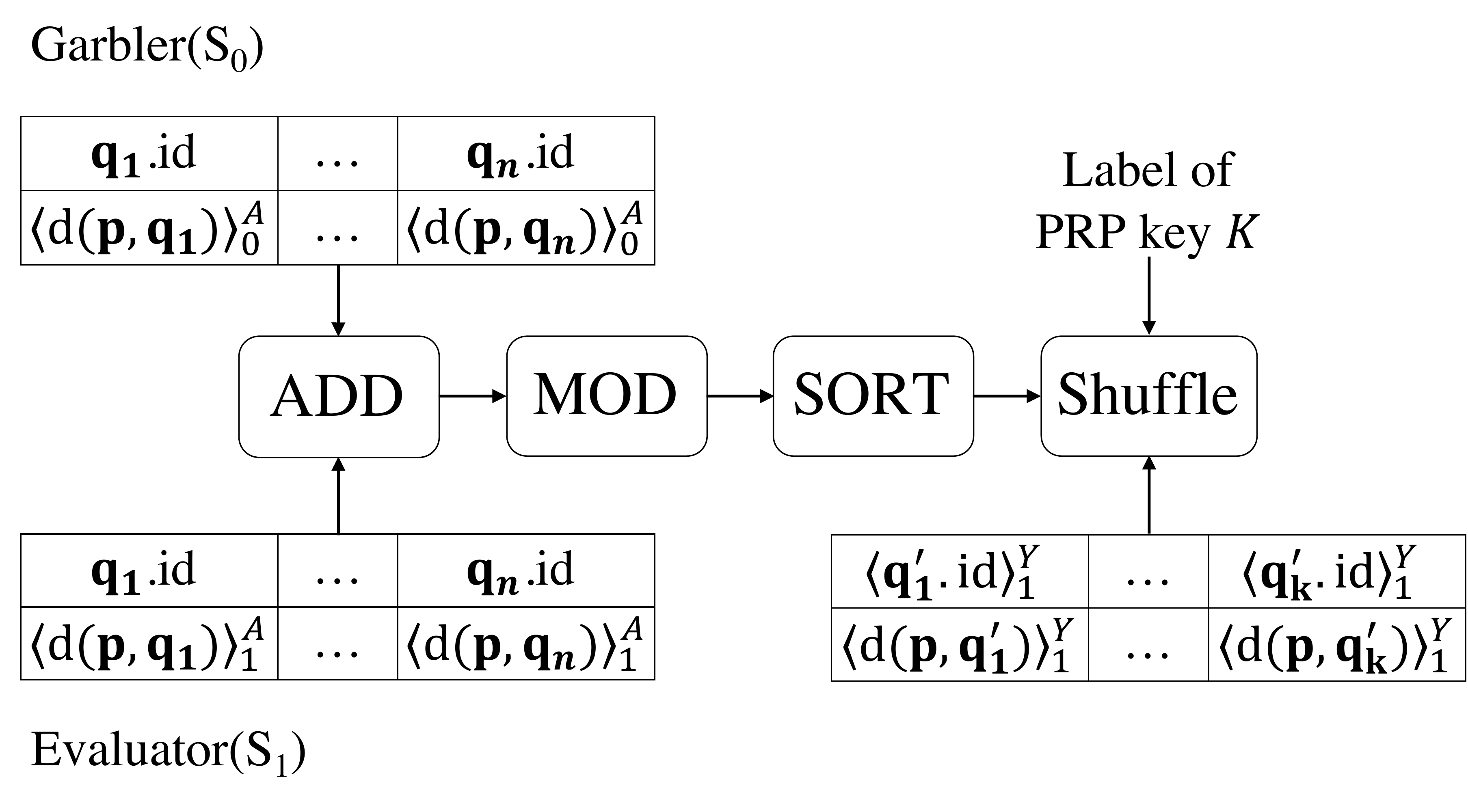}
	}
	\hfill
	\subfloat[Max]{
		\label{fig:maxmin}
		\includegraphics[width=0.46\linewidth]{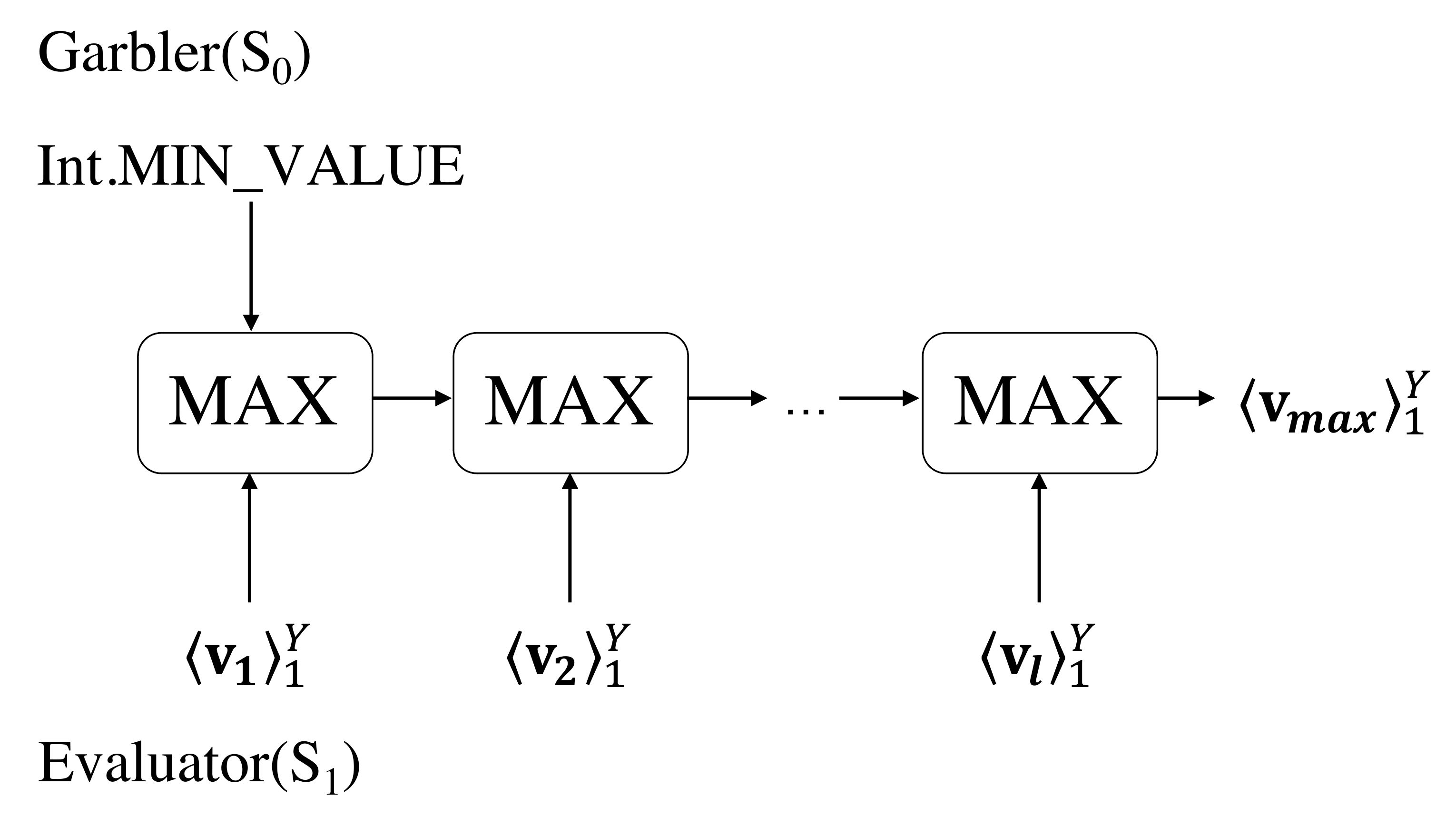}
	}
	
	\subfloat[Randomise]{
		\label{fig:randomise}
		\includegraphics[width=0.5\linewidth]{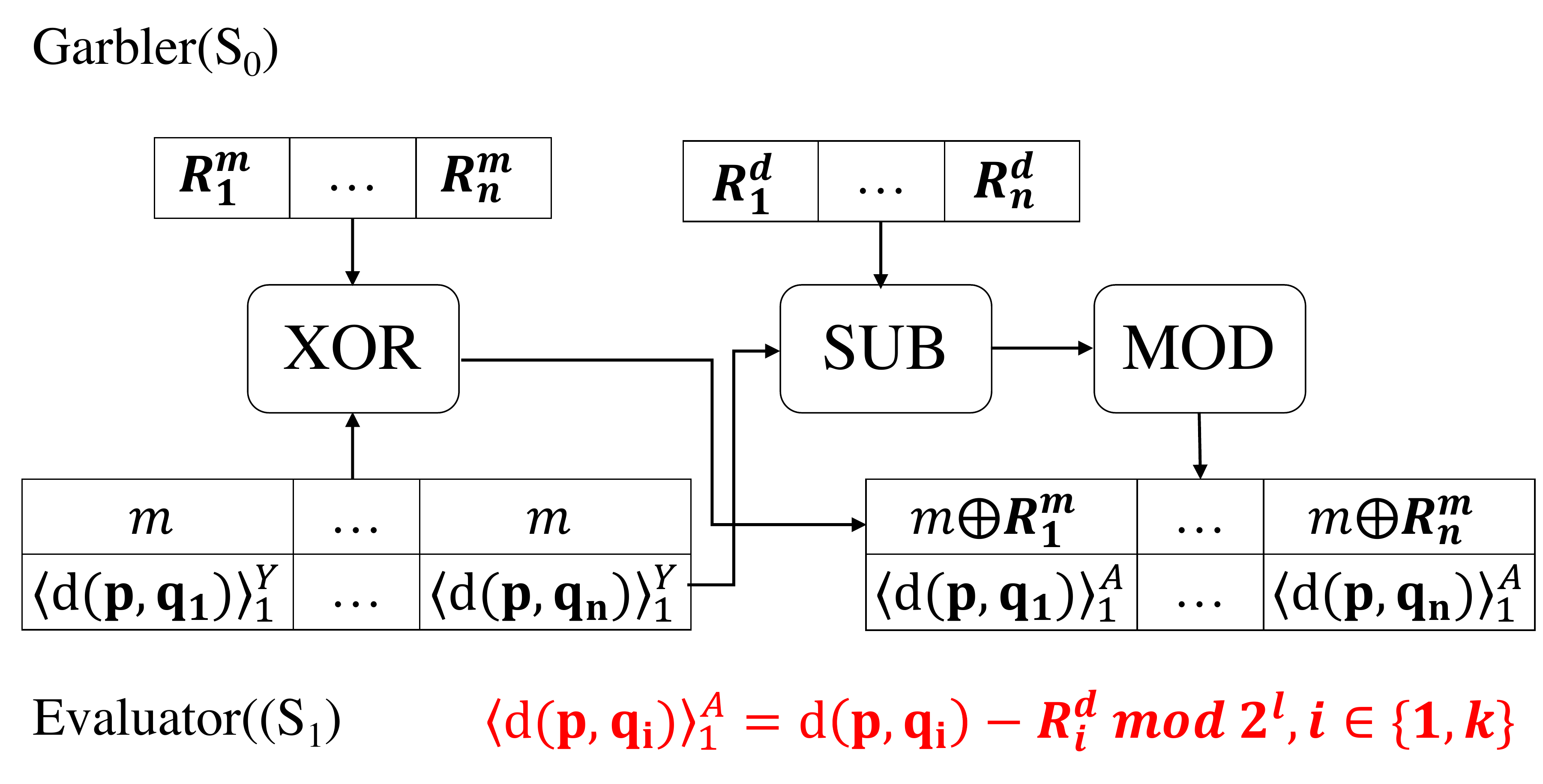}
	}
	\hfill
	\subfloat[Derandomise]{
		\label{fig:derandomise}
		\includegraphics[width=0.46\linewidth]{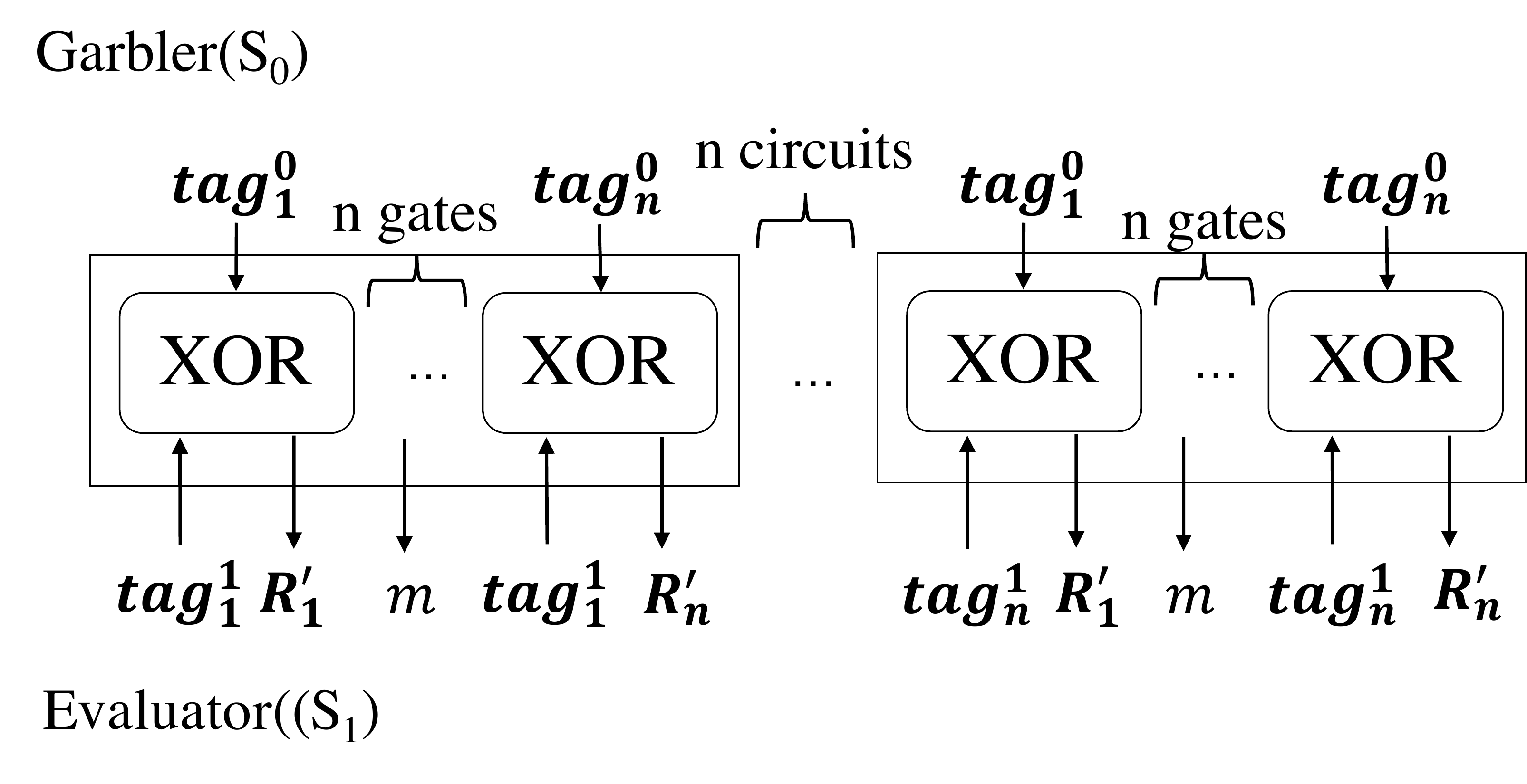}
	}
	\caption{The circuit structure of cryptographic sub-modules.}
	%\vspace{-10pt}
	\label{fig:circuit}
\end{figure*}

{\noindent\bf \textsc{SortShuffle}.} Fig.~\ref{fig:sortshuffle} shows the structure of the secure sorting module for kNN computation, our system follows the standard procedure (see Section~\ref{subsec:secureComp} for details) to evaluate the circuit in Fig.~\ref{fig:sortshuffle} and receives an unordered kNN list as the result.
The secure sorting module inputs the shares of distance to an $\mathsf{A2Y}(\cdot)$ function implemented via an efficient scheme in~\cite{demmler2015aby} to convert the additive shares to Yao's shares.
It then adopts the sorting circuit based on sorting network~\cite{batcher1968sorting} to sort the Yao's shares from the $\mathsf{A2Y}(\cdot)$ function.
In addition, the garbler concatenates the sorting circuit with a shuffling circuit based on a pseudorandom permutation (PRP)~\cite{luby1988construct}, and the evaluator supplies a random key $K$ to disrupt the order of the kNN list and remove the remaining points. 
Finally, the \textsc{SortShuffle} circuit outputs the Yao's shares of the unordered kNN list, which ensures that the order does not reveal in the sorting procedure.

%Given $\Server_0$ as the garbler and $\Server_1$ as the evaluator, both parties input the list of computed distance shares between $\mathbf{p}$ and $\mathbf{q_i}, 1\leq i\leq l$ and the id to run $\textsc{SortShuffle}(L, k)$ protocol as follows: 
%%
%\begin{enumerate}[leftmargin=*]
%	\setlength{\itemsep}{0pt}
%	\setlength{\parsep}{0pt}
%	\setlength{\parskip}{0pt}
%	\item The garbler generates the circuit as shown in Fig.~\ref{fig:sortshuffle} and then gives the circuit to the evaluator.
%	\item The garbler sends the encoded input according to its local shares and point identities to the evaluator. Doing so, it prevents the evaluator from learning the information about data points on the garbler.
%	\item The evaluator runs oblivious transfer (OT) protocol to acquire the encoded input of its local shares and point identities. This prevents the garbler from learning the information about data points on the evaluator.
%	\item The evaluator runs OT protocol to acquire the encoded input of a random key $K$ to evaluate the shuffling gate to protect the order.
%	\item The evaluator uses the given inputs to evaluate the circuit and stores the outputted Yao's share.
%\end{enumerate}

{\noindent\bf \textsc{Max}.} To retrieve the $k$-distance from an unordered kNN list, the \textsc{kDist} algorithm employs the \textsc{Max} circuit shown in Fig.~\ref{fig:maxmin}.
The circuit takes as inputs a list of Yao's shares (e.g., Yao's shares of distances in the kNN list), and it consists of a chain of MAX gates to compute the maximum value (e.g., the $k$-distance in \textsc{kDist}) of the given inputs.
In order to protect the underlying values (the distances), the output of the \textsc{Max} circuit is also in the form of Yao's shares.

{\noindent\bf \textsc{Randomise}/\textsc{Derandomise}.}
Each entity in the kNN list comprises two data types, i.e., the values that indicate the computed distance, and the identities which assist the server to work consistently.
To hide the repeat patterns after kNN and $k$-distance evaluations, we should protect the above two data types when the server converts Yao's shares back to the additive shares for the storage purpose.
We design a $\textsc{Randomise}$ function (see Fig.~\ref{fig:randomise}) to achieve the above goal:
To protect the distance value, the garbler generates a new random value and garbles the circuit for $\mathsf{Y2A}(\cdot)$ to re-share the distance as the additive shares;
To hide the identity on servers, we introduce a flag independent from the data point id to aid the server to find the position of corresponding shares in its counter-party before starting the computations.
More specifically, the evaluator selects a magic number $m$ to de-identify its local points and leverages random numbers $R^m$ generated by the gabler to mask $m$ via xor operations. 
%
%Fig.~\ref{fig:randomise} demonstrates the circuit structure of the %
%The garbler generates two random vectors and employs the $\mathsf{B2Y}(\cdot)$ to re-share the Yao's shares as additive shares again.
%
%Also, it leverages the XoR operation to mask the flag chosen by the evaluator.
%
After circuit evaluation, the garbler stores the generated random vectors as its local data point shares and the evaluator takes the output of the circuit as the new data point shares. 

The $\textsc{Derandomise}$ function is used to pair the randomised shares between two servers. 
As shown in Fig.~\ref{fig:derandomise}, for a randomised list with $n$ elements, the $\textsc{Derandomise}$ function generates $n^2$ xor gates revealing the ``paired'' positions, i.e., the position where the xor gate returns $m$. 
The server then exploits its local shares to run the following secure protocols according to the revealed position.
After computing, two servers run \textsc{Randomise} function again to invalidate the revealed patterns.

\subsection{Data Preprocessing}\label{sec:preprocess}
{\noindent\bf Overview.} Input preprocessing runs for all data points receiving from some DAUs (e.g., sensors).
As shown in Algorithm~\ref{alg:input}, the gateway performs a two-step preprocessing over the received data points before giving them to the server for secure outlier detection.
The first step is to dissolve the input format mismatch between the client and the server. 
Namely, the data point from DAUs consists of fractional numbers and may also include negative numbers, while the additive sharing scheme in our protocol only works over non-negative integers.
Thus, the gateway should pre-process these received coordinates via normalisation and rounding to meet the input requirement of cryptographic primitives before it shares the data to servers
After preprocessing, the gateway generates the additive shares for these adjusted data points and distributes the generated shares to two cloud servers. The detailed construction of the above two preprocessing steps are discussed below:

\begin{algorithm}[!t]
	\caption{Input Preprocessing Phase}
	\label{alg:input}
	\begin{algorithmic}[1]
		\Require Data point set $\mathbf{P}$
		\For{each $\mathbf{p}\in \mathbf{P}$}
		\State $\widetilde{\mathbf{p}}\leftarrow\textsc{Normalise}(\mathbf{p})$
		\State $[\![\mathbf{p}]\!]\leftarrow\textsc{Rounding}(\widetilde{\mathbf{p}}, l_D)$
		\For{i=1 to n}
		\State $\langle [\![p_i]\!]\rangle^A\leftarrow\mathsf{Shr^A}([\![p_i]\!])$
		\EndFor
		\State send $\langle \mathbf{p}\rangle_0=\{\mathbf{p}.id, \langle [\![\mathbf{p}]\!]\rangle^A_0\}$ to $\Party_0$ and $\langle \mathbf{p}\rangle_1=\{\mathbf{p}.id,\langle [\![\mathbf{p}]\!]\rangle^A_1\}$ to $\Party_1$
		\EndFor
	\end{algorithmic}
\end{algorithm}

\noindent 1) \textsc{Normalise}: This function runs to eliminate the negative numbers in coordinates.
For each coordinate, we assume that the maximum/minimum values are fixed at the beginning of data collection, as it is possible for the gateway to know these parameters referring to the hardware specification of DAUs.
Therefore, the gateway can store the maximum/minimum values ($d_i^{max}$ and $d_i^{min}$) for each $i\in[1, n]$.
When the gateway receives a data point ${\bf p}$, it extracts its coordinate value $p_i$ and computes $(p_i-d_i^{min})\cdot(d_i^{max}-d_i^{min})^{-1}$, which outputs a value $\widetilde{p_i}\in[0,1]$ as the corresponding normalised coordinate value for $p_i$.

\noindent 2) \textsc{Rounding}: After normalisation, the coordinates of a normalised point $\widetilde{\bf p}$ have only positive fractional numbers in $[0,1]$.
To handle fractional coordinate values, we introduce a rounding factor $l_D$ to scale up the fractional number into an integer $[\![p_i]\!]=\lfloor\widetilde{p_i} \cdot 2^{l_D}\rfloor$, while preserving $2^{l_D}$ bits in the fractional part of the original number. 
This is a common strategy adopted in several prior works~\cite{bost2015machine,mohassel2017secureml}. As illustrated in the evaluation, the accuracy of the outlier model is not affected under a deliberately selected $l_D$. 

{\noindent\bf Discussion.} The input preprocessing should be applied to all new arrival data points on the gateway before the gateway gives it to the server.
Nevertheless, this will not incur a heavy workload on the gateway and lead to a noticeable delay to the system performance for the following two reasons:
First, the input preprocessing phase can run independently for each data point. Thus, the gateway can leverage parallel processing to handle the received data points in a batch, which can highly improve the preprocessing process.
Besides, the gateway does not involve any computation task other than input preprocessing under the two-server setting. The main computation of the outlier detection algorithm is located on the server.

\begin{algorithm}[!t]
	\caption{Initialisation Phase}
	\label{alg:init}
	\begin{algorithmic}[1]
		\Require Shared data point set $\langle\mathbf{P}\rangle$, Parameter $k, \langle R\rangle^A$
		\Ensure Shared outlier List $\langle O\rangle$
		\For{ each $\langle\mathbf{p}\rangle\in \langle\mathbf{P}\rangle$}
		\State $\langle temp\rangle^Y\leftarrow \textsc{kNN}(\langle\mathbf{P}\rangle, \langle\mathbf{p}\rangle, k)$
		\State $\langle\mathbf{p}\rangle.D\leftarrow\textsc{Randomise}(\langle temp\rangle^Y)$
		\State $\langle dist\rangle^Y \leftarrow \textsc{kDist}(\langle temp\rangle^Y)$
		\State $\langle\mathbf{p}\rangle.D^k\leftarrow \textsc{Y2A}(\langle dist\rangle^Y)$
		\If{$\langle dist\rangle^Y > \mathsf{A2Y}(\langle R\rangle^A)$}
		\State Add $\langle\mathbf{p}\rangle$ into $\langle O\rangle$
		\EndIf
		\EndFor
	\end{algorithmic}
\end{algorithm}
\subsection{Initialisation}\label{sec:init}
{\noindent\bf Overview.} For the first batch of the preprocessed data points (their additive shares) from the gateway, the server invokes the initialisation phase to create the outlier model.
To realise this phase, our system adapts the $D_n^k$ outlier detection algorithm from~\cite{ramaswamy2000efficient} as it can be implemented via arithmetic operations and sorting only, which perfectly suits the secure computation model we used.
In particular, the server uses the received data points and some pre-set parameters to execute the algorithm and gets the $k$-nearest neighbours of each data point as well as the corresponding $k$-distance (denoted as $D^k$).
Consequently, it compares $D^k$ with the distance threshold $R$ to find the outliers (i.e., if $D^k$ is greater than $R$, the point is an outlier).
The server also stores the computed information, i.e., $k$-nearest neighbours, $D^k$ values and the distance threshold $R$ (as additive shares) to support the update phase (see Section~\ref{sec:update}).
The completed procedure of the $D_n^k$-based privacy-preserving outlier detection is shown in Algorithm~\ref{alg:init}. 

{\noindent\bf Discussion.} The initialisation is a time-consuming procedure, as it follows a nested loop (NL) strategy, i.e., it traverses each pair of data points, which infers an $\mathrm{O}(\beta^2)$ computational complexity, where $\beta$ is the number of data points in the batch.
Despite the relatively higher computation cost, we argue that this phase only needs to run once for the entire outlier modelling process, and the model can be updated within $\mathrm{O}(S\cdot W)$ (see Section~\ref{sec:update} for details).

In terms of security, the algorithm with the NL strategy executes the same sequence of operations over all data points.
Hence, the initialisation phase is a data-oblivious process under the two-party secure computation context, that is, the initialisation phase only reveals the information about outliers. 
Conversely, the other information, such as the coordinates, kNN list, and the memory access pattern during the outlier detection process, can be kept in secret.
\subsection{Outlier Query}
{\noindent\bf Overview.} The system user can issue a data point query to check whether the data is an outlier or not by referring to the outlier model on the server side.
The query consists of a preprocessed data point and an outlier threshold.
Once the server receives a query, it evaluates the distance between the query point and outliers using the given additive shares.
Consequently, it utilises the garbled circuit to compare the computed distances and the threshold and produces the final assertion without revealing any sub-result (i.e., which distance is smaller than the threshold).
Algorithm~\ref{alg:query} outlines the query process on each server. Next, we present the detailed construction of the assertion function.

\begin{algorithm}[!t]
	\caption{Query Phase}
	\label{alg:query}
	\begin{algorithmic}[1]
		\Require Shared query point $\langle\mathbf{q}\rangle$ , Parameter $\langle\epsilon\rangle^A$
		\Ensure An assertion (True or False)
		\State $D\leftarrow \{\}$
		\For{ each $\langle\mathbf{o}\rangle\in \langle O\rangle$}
		\State Compute $\langle d\rangle^A\leftarrow \langle \mathsf{d}(\langle\mathbf{q}\rangle, \langle\mathbf{o}\rangle)\rangle^A$ and put $\langle d\rangle^A$ into $D$
		\EndFor
		\State $C\leftarrow \{\}$
		\For{ each $\langle d\rangle^A\in D$}
		\State Compute $\langle r\rangle^Y\leftarrow \mathsf{A2Y}(\langle d\rangle^A)\leq\mathsf{A2Y}(\langle\epsilon\rangle^A)$ and put $\langle r\rangle^Y$ into $C$
		\EndFor
		\State \Return $\textsc{Or}(C)$
	\end{algorithmic}
\end{algorithm}

{\noindent\bf Assertion function.}  In the assertion function , the server makes a comparison between all distances and the given outlier threshold (line 5 -- 8 in Algorithm~\ref{alg:query}). 
If one of those distances is not greater than the threshold, the query point is considered as an outlier, so the server returns `True'; otherwise, it returns `False'. 
Finally, the output assertion is generated via an $\textsc{OR}$ gate, which mixes each pair of distance comparison as the final output.

%%
%After receiving the query from the client, the server adopts the same distance metric as in the initialisation phase to compute the distance between the query point and outliers from the given shares. 
%
%1) Comparison: Then, the server makes a comparison between all distances and the given outlier threshold. If one of those distances is not greater than the threshold, the query point is considered as an outlier, so the server returns 'True'; otherwise, it returns 'False'. 
%%
%During this process, the server should only know the final assertion, but not any intermediate result (e.g., each pair of comparison result) and the input.
%%
%To achieve this requirement, the server firstly converts additive shares to Yao's shares as in the initialisation phase and generates comparison circuits to perform distance check.
%%
%Subsequently, the server attaches an $\textsc{Or}$ gate to mix the comparison results as the final output. 
%%
%Due to the security property of the garbled circuit~\cite{bellare2012foundations}, the server, in this case, only gets the mixture of these comparison results.

{\noindent\bf Discussion.} Query phase is an efficient stage, as it only performs arithmetic operations and comparison with the known outlier list.
Therefore, its computational complexity is bounded by the size of the outlier list $\mathbf{O}$, which is much smaller than the other phases.
In addition, Algorithm~\ref{alg:query} is also a data-oblivious algorithm because it loops for each outlier to produce the result.
During this process, the server only knows the final assertion, but not any intermediate result (e.g., each pair of comparison result) and the input.

\begin{algorithm}[!t]
	\caption{Update Phase}
	\label{alg:update}
	\begin{algorithmic}[1]
		\Require Active point set $\langle \mathbf{P_a}\rangle$, Incoming point set $\langle \mathbf{Q}\rangle$, Parameter $\langle R\rangle^A$
		\Ensure Shared outlier List $\langle \mathbf{O}\rangle$
		\State Remove expired points from $\langle \mathbf{P_a}\rangle$ and $\langle \mathbf{O}\rangle$
		\For{ each $\langle \mathbf{q}\rangle\in \langle \mathbf{Q}\rangle$}
		\State Compute $\langle \mathbf{temp}\rangle^Y\leftarrow \textsc{kNN}(\langle \mathbf{P_a}\rangle, \langle \mathbf{q}\rangle, k)$
		\State $\langle dist\rangle^Y \leftarrow \textsc{kDist}(\langle \mathbf{temp}\rangle^Y)$
		\State $\langle\mathbf{q}\rangle.D^k\leftarrow \textsc{Y2A}(\langle dist\rangle^Y)$
		\If{$\langle dist\rangle^Y >\mathsf{A2Y}(\langle R\rangle^A)$}
		\State Add $\langle \mathbf{q}\rangle$ to $\langle \mathbf{O}\rangle$
		\EndIf
		\State Add $\langle \mathbf{q}\rangle$ to $\langle \mathbf{P_a}\rangle$
		\For { each $\langle \mathbf{t}\rangle^Y\in \langle \mathbf{temp}\rangle^Y$}
		\State Recover $\mathbf{t}.id$ from $\langle \mathbf{t}.id\rangle^Y$ 
		\State Retrieve $\langle \mathbf{a}\rangle$ from $\langle \mathbf{P_a}\rangle$ based on $\mathbf{t}.id$
		\If{$\langle \mathbf{a}\rangle\in\langle \mathbf{O}\rangle$}
		\State $\langle \mathbf{a}\rangle.D\leftarrow \textsc{Derandomise}(\langle \mathbf{a}\rangle.D)$
		\State $\langle \mathbf{a}\rangle.D.\textsc{Add}(\{\mathbf{q}.id, \mathsf{B2A}(\langle\mathbf{t}.d\rangle^Y)\}))$
		\State $\langle temp_a\rangle^Y\leftarrow\textsc{SortShuffle}(\langle \mathbf{a}\rangle.D, k)$
		\State $\langle \mathbf{a}\rangle.D\leftarrow\textsc{Randomise}(\langle temp_a\rangle^Y)$
		\State $\langle \mathbf{a}\rangle.D^k\leftarrow \mathsf{Y2A}(\textsc{kDist}(\langle temp_a\rangle^Y))$
		\If{$\textsc{kDist}(\langle temp_a\rangle^Y)\leq \mathsf{A2Y}(\langle R\rangle^A)$}
		\State Remove $\langle \mathbf{a}\rangle$ from $\langle \mathbf{O}\rangle$
		\EndIf
		\EndIf
		\EndFor
		\EndFor
	\end{algorithmic}
\end{algorithm}
\subsection{Model Update}\label{sec:update}
{\noindent\bf Overview.} In the model update phase, each server receives a new batch of the preprocessed data points and computes a new outlier model, which takes these new data points into consideration.
To ensure the efficiency of this phase, the update protocol for PPOD uses the sliding window model and maintains a list of active points and only recomputes/reports the outliers for the active points.
Also, the update algorithm only updates the data points that affect by the added/expired data points, which is consistent to the incremental algorithms~\cite{angiulli2007detecting,kontaki2011continuous} for the plaintext outlier detection scheme.
%It then adapts an update protocol based on exact-Storm~\cite{angiulli2007detecting}, which works on the sliding window to update the model.
%
In particular, the update protocol removes the expired points from the active point set $\mathbf{P_a}$ and the outlier list $\mathbf{O}$.
Then, it computes the kNN and $k$-distance information for the new data point by utilising the remaining $\mathbf{P_a}$ and determines whether the new point is an outlier.
Later, the protocol updates the points in $\mathbf{P_a}$ which are also in the kNN list of the new data point.
%
%Specifically, it first integrates the update function in Section~\ref{sec:knn} to update the kNN list and then uses the new kNN list to update the $k$-distance and outlier list.
%
The procedure of the update phase is given in Algorithm~\ref{alg:update}.

{\noindent\bf Discussion.} The simplest solution is to run the initialisation protocol for the updated dataset. However, as mentioned in Section~\ref{sec:init}, the initialisation is an inefficient phase ($\mathrm{O}(|\mathbf{P}|^2)$, where $|\mathbf{P}|$ is the size of dataset).
%
%Moreover, $|\mathbf{P}|$ is continually increasing with the new data point arrivals; the update phase is also continually slowing down if it relies on the initialisation algorithm to update.
%
Compared to the naive approach in the above, the complexity of the proposed update approach is lower:
For each new data point, the update phase only refers the active data points to compute the kNN list, which takes $\mathrm{O}(W)$, where $W$ is the sliding window size, and $\mathrm{O}(k)$ to update the existing information.
And the whole update procedure runs for the new points after sliding (add $S$ new points), which indicates that the overall runtime complexity is $\mathrm{O}(S\cdot W)$

In terms of the security, the update approach does not guarantee the data-oblivious, because it retrieves the id of the kNN list when it updates for the existing data points (line 11 -- 12 in Algorithm~\ref{alg:update}).
Nevertheless, we stress that this is the only additional leakage comparing with the other phases, and it enables a more efficient update phase.

\section{Security Analysis}\label{sec:security}
We give the security analysis following the classic paradigm of comparing the real-world execution of the protocol to an ideal-world execution where a trusted third party evaluates the functions on behalf of the involved parties.
The only difference is that we consider an ideal world that the adversary is allowed to learn the $k$-nearest neighbours of a new arrival data point when adding it into the sliding window.
Note that we leverage an OT-hybrid model where parties are given access to the trusted party computing the ideal function of OT. The following theorem shows that the PPOD protocol is secure with the given leakage function in this hybrid model. 
Thus, the PPOD protocol remains secure if the trusted party is replaced by the real OT.

\begin{algorithm}[!t]
	\caption{Ideal Function $\mathcal{F}_k$}
	\label{alg:idealk}
	{\bf Parameters:} Client $\Client$ and servers $\Server_0$, $\Server_1$.
	
	{\bf Input:} On input $\langle \mathbf{p}\rangle_i, i\in\{0,1\}$ from $\Client$, stores it locally.
	
	{\bf $\text{kNN}_{id}$:} On input the query point $\langle \mathbf{q}\rangle_i, i\in\{0,1\}$ and the shared point set $\langle \mathbf{P}\rangle_i$ from $\Server_i$, the functionality returns an unordered point list with id only.
	
	{\bf $\text{kNN}_{dist}$:} On input the query point $\langle \mathbf{q}\rangle_i, i\in\{0,1\}$ and the shared point set $\langle \mathbf{P}\rangle_i$ from $\Server_i$, the functionality returns an unordered list of the shared distances only.
	
	{\bf kDist:} On input the query point $\langle \mathbf{q}\rangle_i, i\in\{0,1\}$ and the shared point set $\langle \mathbf{P}\rangle_i$ from $\Server_i$, the functionality returns the shared distance between the query point and its $k^{th}$ nearest neighbour.
	
	{\bf Update}: On input the kNN list $\langle \mathbf{p}\rangle_i.D$ of point $p$ and a shared points $\langle \mathbf{q}\rangle_i$ with the shared distance from $\Server_i$, the functionality updates $\langle \mathbf{p}\rangle_i.D$ and returns an unordered list of the shared distances only.
\end{algorithm}
To start with, we give a security analysis for the secure kNN and $k$-distance modules in Section~\ref{sec:knn}, as our PPOD protocol highly depends on these modules. 
\begin{theorem}\label{thm:kNN}
	Consider a protocol where clients distribute shares of data points among two servers who run our PPOD protocol from Section~\ref{sec:construction}. 
	In the OT-hybrid model, the protocol $\Pi_k$ realises the ideal function $\mathcal{F}_k$ in Algorithm~\ref{alg:idealk} in presence of semi-honest but non-colluding adversaries.
\end{theorem}

{\noindent\bf Proof.} We denote the secure kNN and $k$-distance protocols as $\Pi_k$, and our proof shows that $\Pi_k$ securely realises the ideal functions $\mathcal{F}_k$ in Algorithm~\ref{alg:idealk}. As the adversary in our model only corrupts one server at most, and the view of two servers are slightly different (one garbler and one evaluator), we separately consider the scenario that the adversary $\Adv_i, i\in\{0,1\}$ corrupts $\Server_i$.
For each $\Adv_i$, we describe how to construct a simulator $\Sim_i$ that simulates $\Adv_i$ in the ideal model.
For two varieties of the kNN evaluations (i.e., $\text{kNN}_{id}$ and $\text{kNN}_{dist}$), the only difference between them is how they handle the output of \textsc{SortShuffle}.
In particular, in $\text{kNN}_{id}$, $\Sim_0$ returns the identities from the trusted party to $\Adv_0$ and $\Sim_1$ should give the simulated decoded information of identities to $\Adv_1$.
On the other hand, in $\text{kNN}_{dist}$, both simulators are only required to return the random shares of distance to the adversary.

We claim that $\Adv_i$'s view in the real and ideal model is indistinguishable for the kNN evaluations: 
Since the security of the additive sharing scheme and multiplication triplets ensure the randomness of distance shares, and the protocol is a composition of a sequence of secure modules (\textsc{SortShuffle}, \textsc{Randomise}).
It follows from the modular composition theorem~\cite{canetti2000security} that the adversaries' views are both identical. 
The kDist function is almost identical to the kNN functions except it connects the output of \textsc{SortShuffle} gate to a \textsc{Max} gate to retrieve the maximum distance in kNN list.
Therefore, we can follow the same path to show the security of the kDist function, i.e., the modular composition theorem is applied for \textsc{SortShuffle} gate, \textsc{Max} gate, and \textsc{Y2A} gate to get the same view in real/ideal models.
The update function only involves garbled circuit evaluation, and the security of the garbled circuit ensures that no adversary can learn the input (i.e., previous kNN list) from the output and the execution on the circuit. \hfill\IEEEQED

\begin{algorithm}[h]
	\caption{Ideal Function $\mathcal{F}_o$}
	\label{alg:ideal}
	{\bf Parameters:} Client $\Client$ and servers $\Server_0$, $\Server_1$.
	
	{\bf Input:} On input $\langle \mathbf{p}\rangle_i, i\in\{0,1\}$ from $\Client$, stores it locally.
	
	{\bf Initialise:} On input the first batch of shared points $\langle \mathbf{P}\rangle_i, i\in\{0,1\}$ from $\Server_i$, the functionality initialises the shared outlier list $\langle \mathbf{O}\rangle_i$.
	
	{\bf Query:} On input the shared query point $\langle \mathbf{q}\rangle$ from $\Client$, and the shared outlier list $\langle \mathbf{O}\rangle_i$ from $\Server_i$, the functionality returns 'True' or 'False' to indicate the query point is an outlier or not.
	
	{\bf Update}: On input the active shared points $\langle \mathbf{P_a}\rangle_i$ and the new batch of shared points $\langle \mathbf{Q}\rangle_i$ from $\Server_i$, the functionality updates the shared outlier list $\langle \mathbf{O}\rangle_i$. Besides, it returns the identities of kNN of new arrival data point  $\langle \mathbf{q}\rangle_i\in \langle \mathbf{Q}\rangle_i$ sequentially.
\end{algorithm}

\noindent We now provide the security proof of the PPOD protocol. 
The ideal function of our PPOD is given in Algorithm~\ref{alg:ideal}. The following theorem demonstrates the PPOD scheme is secure under the non-colluding semi-honest server model.
\begin{theorem}\label{thm:outlier}
	Consider a protocol where clients distribute shares of data points among two servers who run the PPOD protocol in Section~\ref{sec:construction}. 
	In the ($\mathcal{F}_k$, OT)-hybrid model, the PPOD protocol adopts the ideal function $\mathcal{F}_o$ with leakage consisting of $k$-nearest neighbours of new arrival data points in Algorithm~\ref{alg:ideal} in semi-honest but non-colluding adversarial model.
\end{theorem}

{\noindent\bf Proof.} We follow the same setting to prove the security of the PPOD system.
In the initialisation phase, $\Sim_i$ runs $\Adv_i$ and sends randomly generated shares in $\mathbb{Z}_{2^l}$ with identity as the shared points to $\Adv_i$. 
Besides, the computation and randomisation of kNN list and $k$-distance can be simulated by calling the ideal function $\mathcal{F}_k.\text{kNN}_{dist}$ and $\mathcal{F}_k.\text{kDist}$.
Finally, $\Sim_0$ utilises a dummy circuit and simulates input labels and plays the role of the trusted server to send the detected outlier identities to simulate the view of $\Adv_0$. 
On the other hand, $\Sim_1$ relies on $\Pi_k$ to get the comparison result between $k$-distance and threshold $R$ and sends the simulated circuit with the same output to $\Adv_1$ as its view.

Now, we illustrate the security of PPOD in each phase, respectively:
During initialisation, $\Adv_i$'s view in the real and ideal model is indistinguishable: $\Sim_i$ provides the random value as the shared points and simulates the garbled circuit via the output of $\mathcal{F}_k$ for the corresponding $\Adv_i$. 
Besides, it uses the ideal function $\mathcal{F}_o$ to return the result to $\Adv_i$.

For the query phase, the simulator leverages the random input to simulate the query points, and then, it can simulate the adversaries' view similarly as above.
In particular, the distance shares is also a random number as it leverages the randomly generated multiplication triplets. 
Moreover, the simulator utilises the simulator of garble circuit to simulate the rest of the protocol and returns the assertion to the client. 
Therefore, the modular composition theorem also implies that the query protocol remains secure after combining the additive sharing scheme and the garbled circuit.

The update phase is almost identical to the initialisation phase, except that it additionally reveals the kNN list of new arrival data, and there is an extra round to update the information of these $k$-nearest neighbours.
Specifically, the update phase requires to call $\mathcal{F}_k.\text{kNN}_{id}$ and updates the points with the returned id.
As a result, the update phase is secure with one extra leakage, as it is the composition of the initialisation phase and the functionalities in $\mathcal{F}_k.\text{kNN}_{id}$ and $\mathcal{F}_k.\text{Update}$.
As $\Pi_k$ securely realises $\mathcal{F}_k$, the PPOD scheme also securely realises the $\mathcal{F}_o$ with the leakage of $k$-nearest neighbours of a new arrival data point in the ($\mathcal{F}_k$, OT)-hybrid model.\hfill\IEEEQED

\begin{figure*}[!t]
	\begin{minipage}[b]{0.33\textwidth}
    	\subfloat[kNN]{
    		\label{fig:kinsert}
	    	\includegraphics[width=\textwidth]{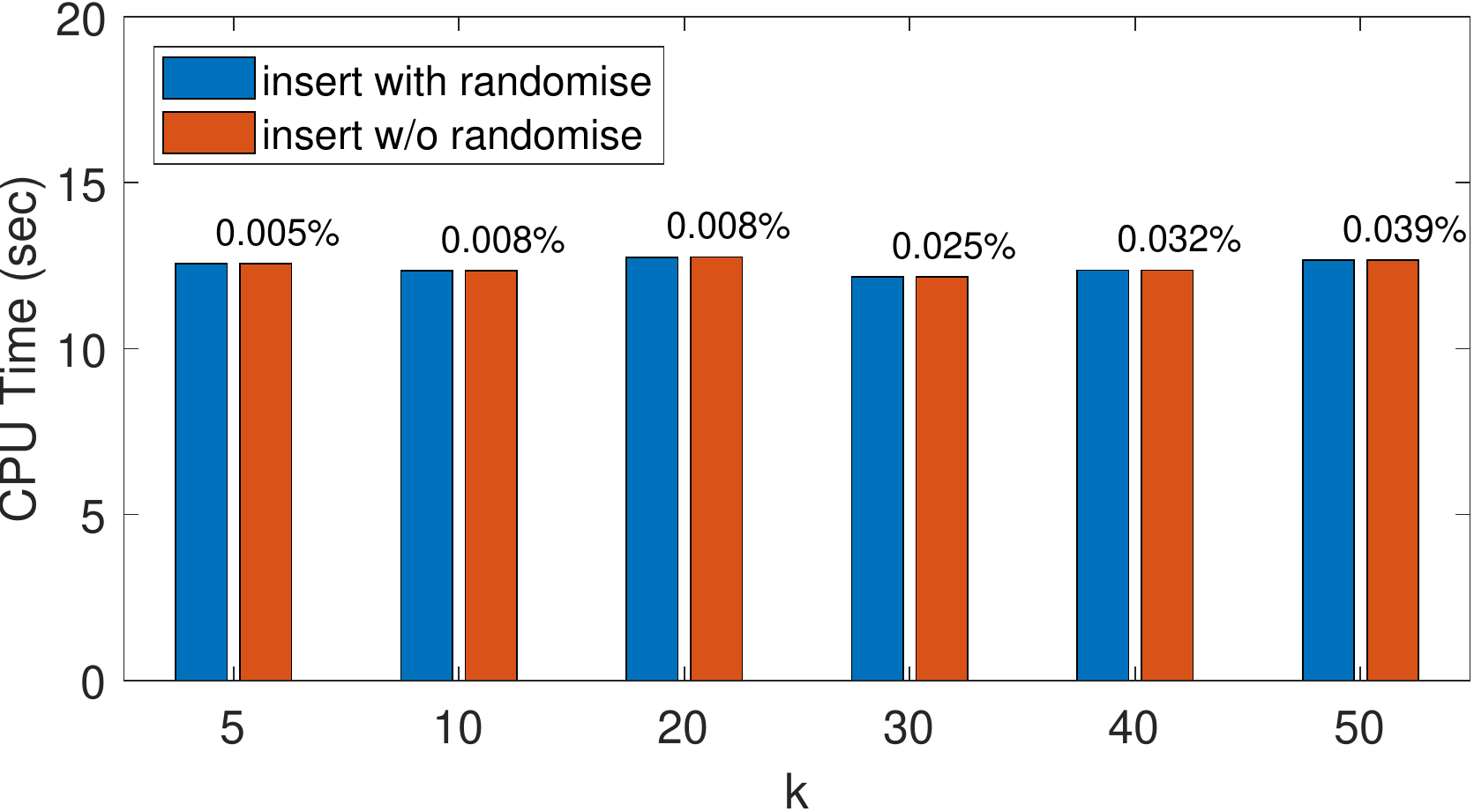}
	    }\\
	    \subfloat[Update]{
	    	\label{fig:kupdate}
	    	\includegraphics[width=\textwidth]{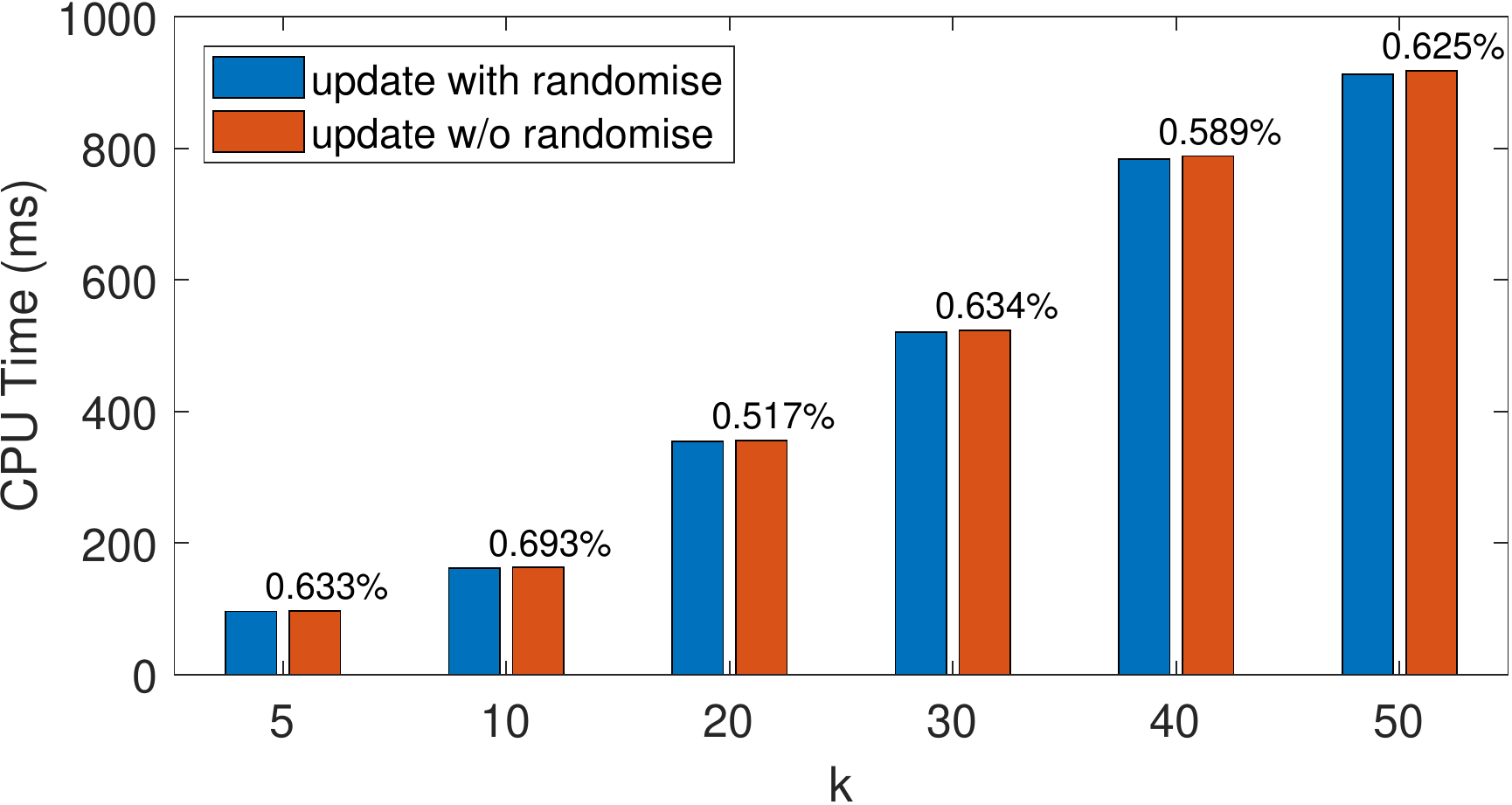}
	    }
	    \label{fig:kNN}
	    \caption{CPU Time of the proposed secure kNN module when varying $k$. Numbers on top of the bars demonstrate the overhead ratio between red and blue bars.}
    \end{minipage}
    \hspace{1pt}
    \begin{minipage}[b]{0.33\textwidth}
    	\subfloat[kNN]{
    		\label{fig:kinsert_memory}
	    	\includegraphics[width=\textwidth]{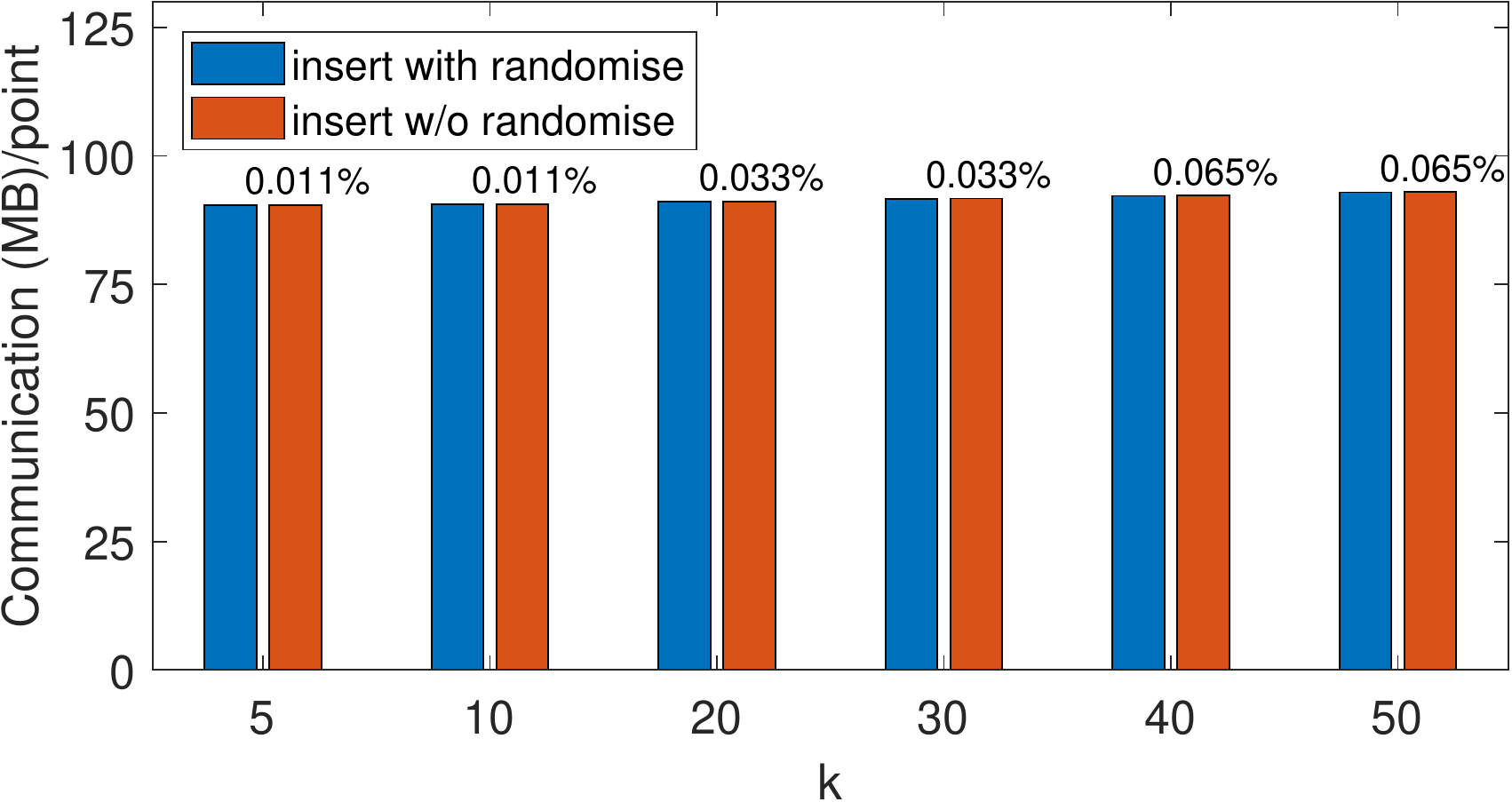}
	    }\\
	    \subfloat[Update]{
	    	\label{fig:kupdate_memory}
	    	\includegraphics[width=\textwidth]{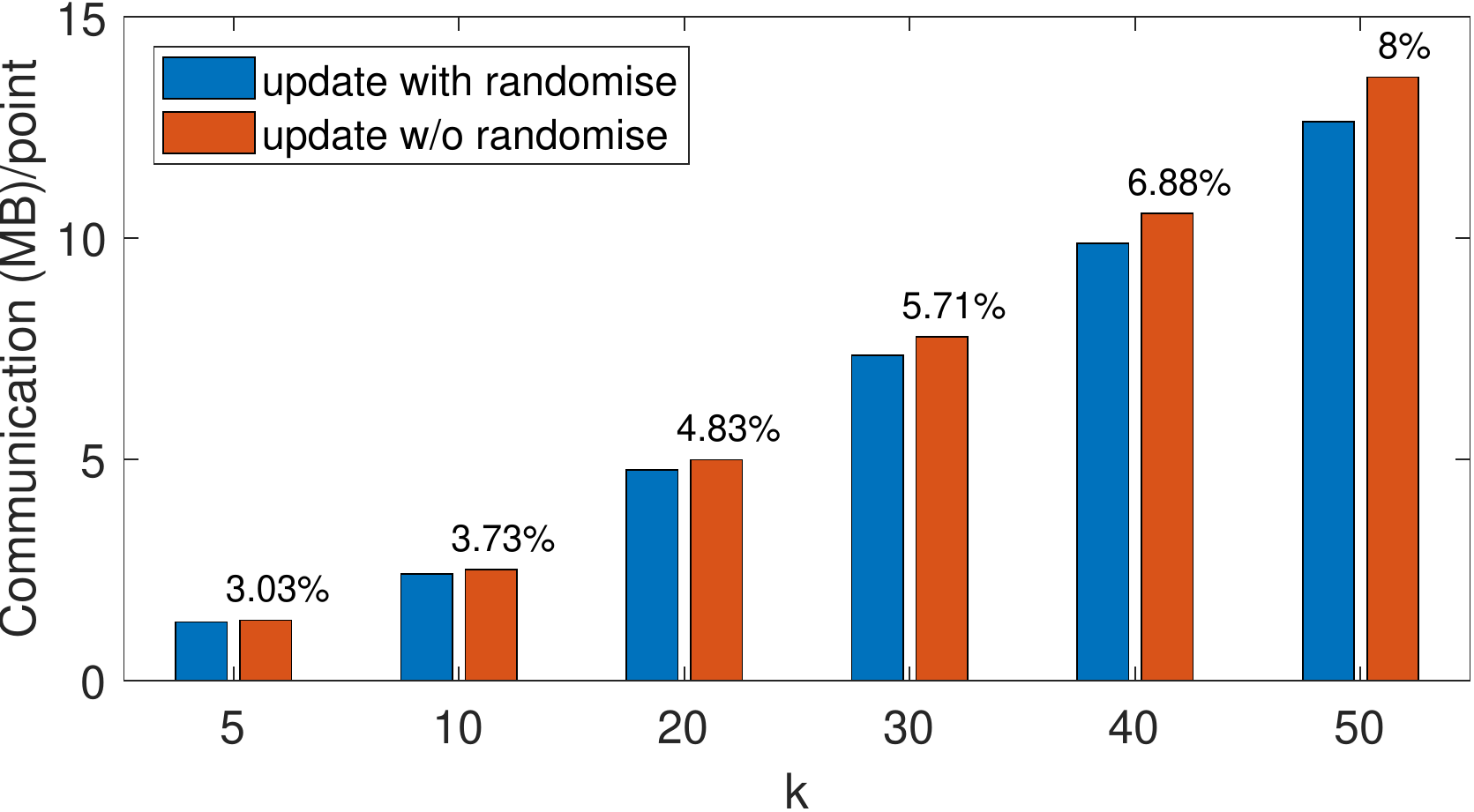}
	    }
	    \label{fig:kNN_memory}
	    \caption{Communication overhead of the proposed secure kNN module with different $k$.Numbers on top of the bars demonstrate the overhead ratio between red and blue bars.}
    \end{minipage}
    \hspace{1pt}
   	\begin{minipage}[b]{0.33\textwidth}
    	\subfloat[Runtime]{
    		\label{fig:wruntime}
	    	\includegraphics[width=\textwidth]{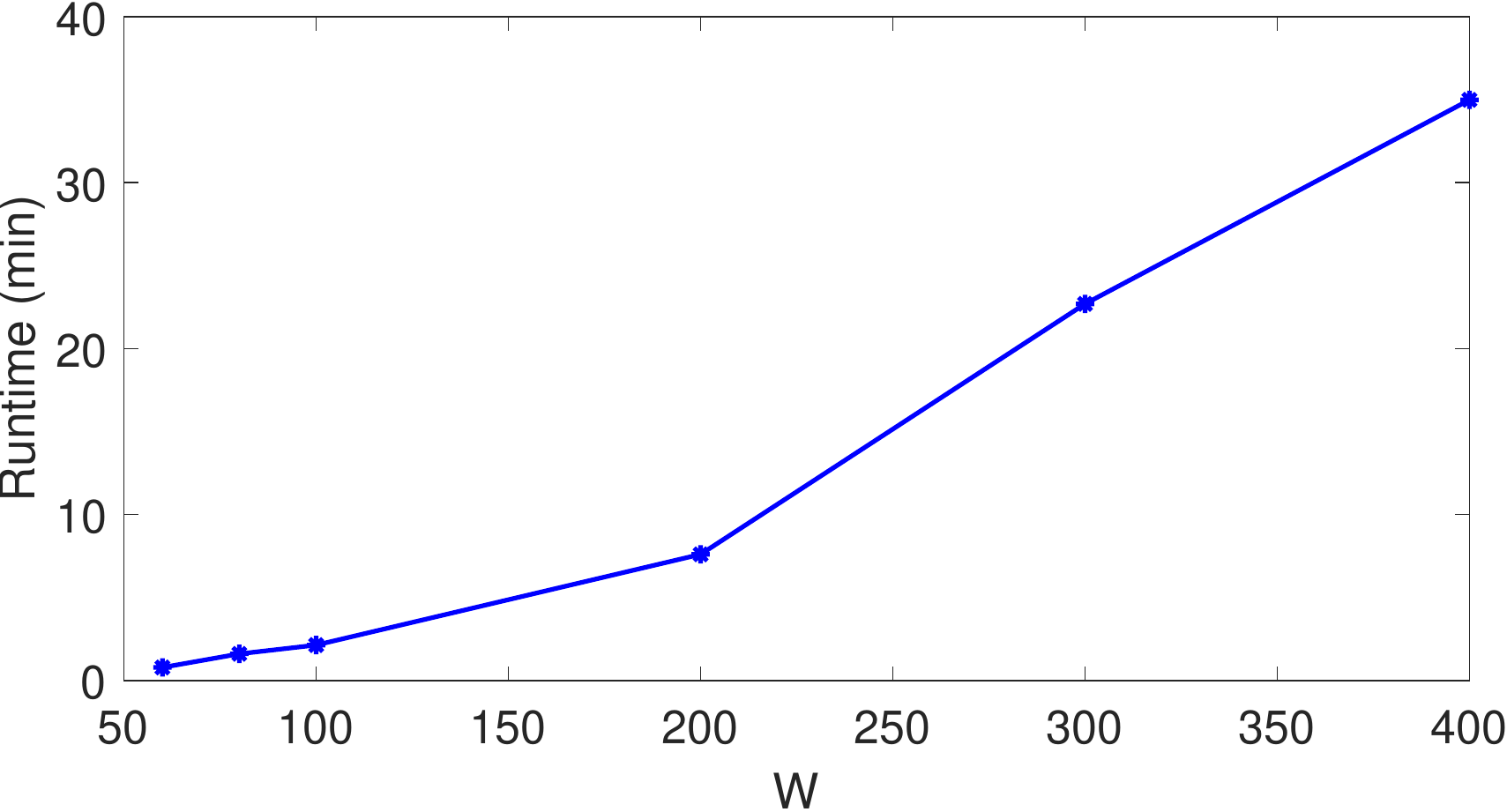}
	    }\\
	    \subfloat[Peak memory]{
	    	\label{fig:wmemory}
	    	\includegraphics[width=\textwidth]{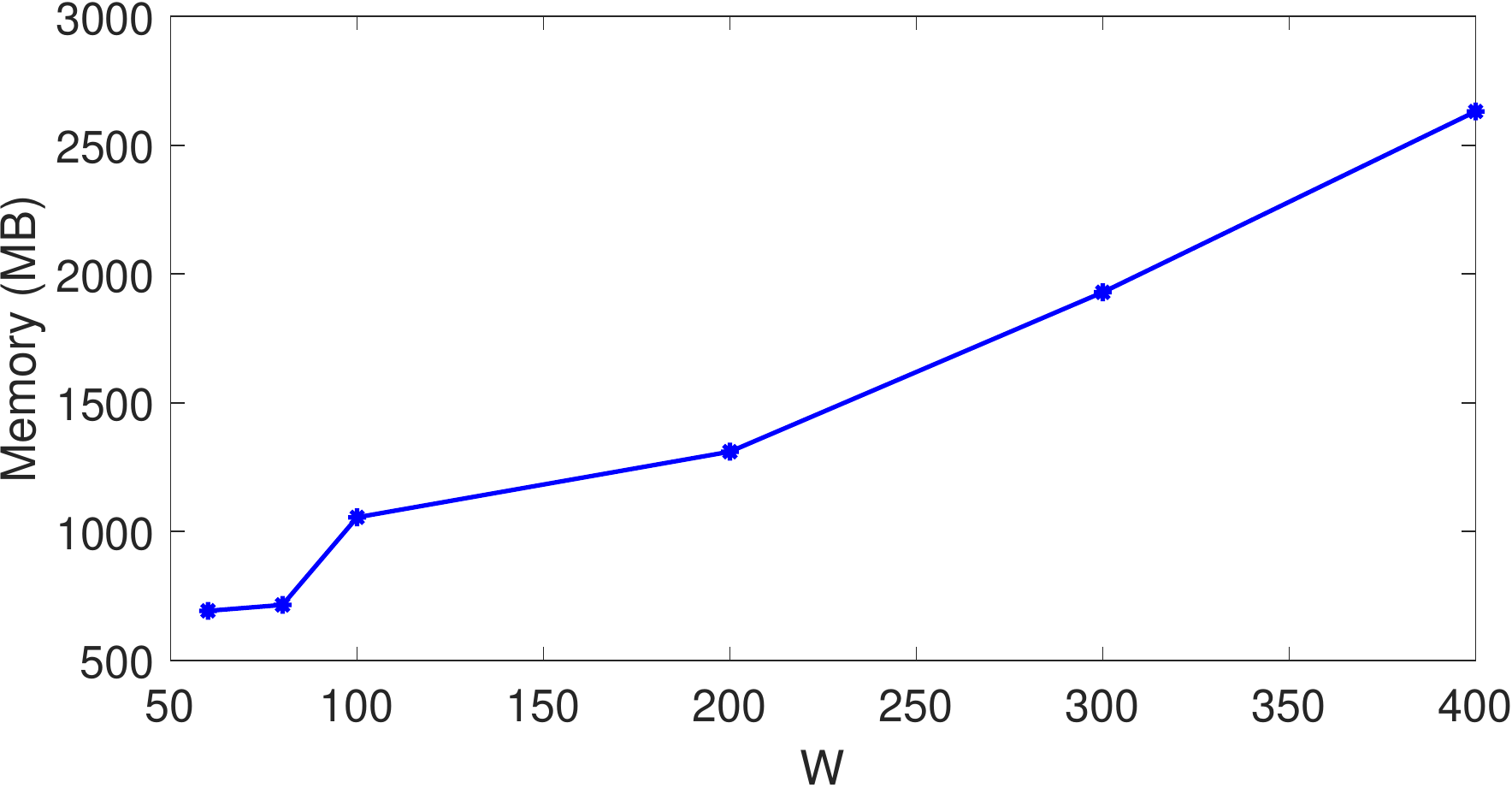}
	    }
	    \label{fig:overall_memory}
	    \caption{The runtime and memory usage of the initialisation phase with varying $W$.}
    \end{minipage}
    \vspace{-15pt}
\end{figure*}

\section{Evaluation}\label{sec:evaluation}
{\noindent\bf Implementation.} We implement our PPOD system in Java. 
To enable the efficient and secure two-party computation on the cloud server, we first implement the additive sharing scheme.
The arithmetic operations in the additive sharing scheme are computed by several regular addition and multiplication operations with the modulo operation over Java primitive types. 
Note that the modulo operation implemented via Java primitive types (e.g. {\bf long}, {\bf int}) is much faster than the native modulo operation in Java {\bf BigInteger} type (about 50x faster).
For the oblivious transfer (OT) and garbled circuit protocol, we leverage FlexSC~\cite{xiao2018gc}, which includes the implementation of extended OTs~\cite{asharov2013more} and the optimised garbled circuit scheme.
To improve the runtime performance of our prototype, PPOD system maintains a pool of pre-computed multiplication triplets, and it periodically refreshes it to avoid extra computation/communication cost on-the-fly.

{\noindent\bf Setup.} The experiments are executed on two EC2 c5.4xlarge instances running Ubuntu 18.04LTS.
Each instance has 16 cores and 48 GB of memory.
Besides, we create a c5.large instance (4 cores and 8GB memory) serving as the client (i.e., gateway) in the PPOD system. It preprocesses and distributes the dataset to the above two more powerful servers to execute the PPOD protocol.
Our servers are connected with a 10Gb NIC.
To evaluate the performance of PPOD, we use a real-world dataset from UCI~\cite{Dua:2019}, which contains $4,200$ records of $16$-dimension. 

{\noindent\bf Parameters.} There are four parameters in our PPOD system: the window size $W$, the slide size $S$, the count threshold $k$ and the distance threshold $R$.
We evaluate the PPOD system under different $W$ and $k$ because they are the main factors affecting the performance of our PPOD.
In particular, $W$ determines the number of distance measurement functions to be executed as well as the input size of the \textsc{SortShuffle} circuit.
On the other hand, $k$ determines the size of \textsc{Randomise}/\textsc{Derandomise} and \textsc{kDist} function, which are frequently used during the update phase.
By default, we set $W=400$, $S=20$, $k=50$ and $R=25,000$ in our dataset.
Unless specified otherwise, all the parameters take on their default values in the experiments.

In the rest of this section, we first benchmark the performance of the kNN module, and then we report the runtime performance of our PPOD. 

\subsection{Performance of the kNN module}
{\noindent\bf CPU Time.} Fig.~\ref{fig:kNN} depicts the resulting CPU time of the secure kNN module in different phases.
In particular, Fig.~\ref{fig:kinsert} shows the CPU time when adding a new point into the current model: despite the increasing of $k$, the CPU time of adding a new point is a constant (around 12s). 
This is because the kNN is executed during the initialisation phase and the update phase to process the new arrival points, and it involves the distance measurement computation and \textsc{SortShuffle} evaluation with the existing data points (380-400 points). 
Compared to the above two steps, the remaining steps, i.e., computing the $k$-distance and \textsc{Randomise} with $k$ inputs, can be done efficiently (less than $17$ ms according to our evaluation).

The CPU time of updating (see Fig.~\ref{fig:kupdate}) an existing point is varying from $96$ ms to $912$ ms with the increasing of $k$.
The update function of the kNN module only runs in the update phase to update the kNN list of the target point.
The parameter $k$ affects the runtime performance of the update function, since the parameter determines the size of kNN lists, and the server takes more time to evaluate a larger circuit to update if the size of kNN lists is larger.
Finally, we examine the impact of the proposed \textsc{Randomise}/\textsc{Derandomise} cryptographic modules. 
As shown in Fig.~\ref{fig:kNN}, the costs of using these two modules are almost negligible (\textsc{Randomise}: 0.6 - 5.7 ms, \textsc{Deandomise}: 0.03 - 1.28 ms), because they only include simple circuit structure (i.e., Subtract gates and free xor gates).
Therefore, these two modules help our PPOD to achieve a better security guarantee with a small cost when computing the kNN . 

{\noindent\bf Communication.} Fig.~\ref{fig:kNN_memory} demonstrates the communication overhead of processing one data point via the kNN module.
It shows a similar pattern as in the CPU time evaluation.
Specifically, the garbler in the kNN module requires to send a constant size of the input (90 MB) to the evaluator, because the major part of the input is the \textsc{SortShuffle} circuit, and its size is dependent on $k$.
The communication overhead of the update function is relatively small (1 MB - 12 MB), but it is proportional to $k$ for the same reason as in the CPU time evaluation, i.e., the generated circuit size is proportional to $k$.
The communication overhead slightly increases when the system facilities the \textsc{Randomise}/\text{Derandomise} cryptographic modules to enhance the security of data points, especially for the \textsc{Derandomise} module, where the size complexity is $\mathrm{O}(k^2)$.
As shown in Fig.~\ref{fig:kupdate_memory}, it incurs at most $8\%$ more communication overhead when the randomisation is deployed.
Nevertheless, we claim that this overhead is affordable, as it only consists of xor gates, which is a small object comparing to the sorting circuit and it is easy to evaluate (free xor gates).

\begin{table}[!t]
	\centering
	\caption{Runtime performance of the PPOD system under default parameters.}
	\label{tlb:runtime}
	\begin{tabular}{|c|c|c|c|c|}
		\hline
		Phase & Preprocess & Initialisation & Query & Update \\ 
		\hline
		Time & 46 ms & 35 min & 217 ms & 9 s \\
		\hline
	\end{tabular}
	%\vspace{-5pt}
\end{table}
\subsection{Performance of PPOD}
First, we note that our proposed PPOD achieves the same accuracy as running the plaintext outlier detection protocol~\cite{angiulli2007detecting} on the unencrypted dataset.
Next, we illustrate the run-time performance of each phase of PPOD in Table~\ref{tlb:runtime}.
It shows that the preprocess and query can be done in several milliseconds, which indicates that the client (the gateway) can preprocess the data point with small computational resources and get a real-time query result regarding the current outlier model.
In addition, although the initialisation needs $35$ minutes to execute, it only runs for the first $400$ data points. 
After initialisation, the system can update the existing point only in $9$ s, which is a moderate runtime in the application context.

{\noindent\bf Impact of $W$.} We further examine the runtime performance and memory usage of the initialisation phase for different $W$s, as this phase highly depends on the window size $W$.
Fig.~\ref{fig:overall_memory} depicts the result runtime and memory usage respectively.
When $W$ increases, the CPU time and memory consumption are expected to increase as well.
Besides, we observe that the memory consumption increases sharply when the $W$ reaches $100$ (see Fig.~\ref{fig:wmemory}).
The increase of $W$ not only affects the size of the generated circuit and the number of multiplication triplets but also the delay of evaluating the circuit and computing distance via triplets.
Therefore, there are more objects residing in the memory for computation, and it leads to the rapid growth of memory consumption.
However, such a memory consumption is in an acceptable level in our evaluation platform (48 GB memory) and the other public clouds such as Azure.

\section{Conclusion}\label{sec:conclusion}
This paper presents a privacy-preserving outlier detection (PPOD) protocol targeting the encrypted incremental dataset.
Our PPOD protocol leverages the advanced cryptographic primitives (i.e., secure two-party computation protocols) to build several secure and efficient modules.
In addition, it adopts the sliding window technique to ensure a practical performance during the update phase with new arrival data points.
We implemented our PPOD as a prototype system, and we provided a performance evaluation based on a real-world dataset to demonstrates its accuracy and efficiency.

\section*{Acknowledgement}
This work was supported in part by the Monash FIT Multidisciplinary Seed Funding Scheme, the Data61 Collaborative Research Project (UbiSENSE for Cities), and an AWS Research Grant.

\bibliographystyle{IEEEtran}
\bibliography{reference}

\begin{thebibliography}{10}
\providecommand{\url}[1]{#1}
\csname url@samestyle\endcsname
\providecommand{\newblock}{\relax}
\providecommand{\bibinfo}[2]{#2}
\providecommand{\BIBentrySTDinterwordspacing}{\spaceskip=0pt\relax}
\providecommand{\BIBentryALTinterwordstretchfactor}{4}
\providecommand{\BIBentryALTinterwordspacing}{\spaceskip=\fontdimen2\font plus
\BIBentryALTinterwordstretchfactor\fontdimen3\font minus
  \fontdimen4\font\relax}
\providecommand{\BIBforeignlanguage}[2]{{%
\expandafter\ifx\csname l@#1\endcsname\relax
\typeout{** WARNING: IEEEtran.bst: No hyphenation pattern has been}%
\typeout{** loaded for the language `#1'. Using the pattern for}%
\typeout{** the default language instead.}%
\else
\language=\csname l@#1\endcsname
\fi
#2}}
\providecommand{\BIBdecl}{\relax}
\BIBdecl

\bibitem{sadik2014research}
S.~Sadik and L.~Gruenwald, ``{Research Issues in Outlier Detection for Data
  Streams},'' \emph{ACM SIGKDD Explorations Newsletter}, vol.~15, no.~1, pp.
  33--40, 2014.

\bibitem{yuan2016privacy}
X.~Yuan, X.~Wang, J.~Lin, and C.~Wang, ``{Privacy-Preserving Deep Packet
  Inspection in Outsourced Middleboxes},'' in \emph{{IEEE INFOCOM}'16}, 2016.

\bibitem{chandola2009anomaly}
V.~Chandola, A.~Banerjee, and V.~Kumar, ``{Anomaly Detection: A Survey},''
  \emph{ACM Computing Surveys (CSUR)}, vol.~41, no.~3, 2009.

\bibitem{fu2018risks}
K.~Fu and W.~Xu, ``{Risks of Trusting the Physics of Sensors},''
  \emph{Communications of the ACM}, vol.~61, no.~2, pp. 20--23, 2018.

\bibitem{salehi2016fast}
M.~Salehi, C.~Leckie, J.~C. Bezdek, T.~Vaithianathan, and X.~Zhang, ``{Fast
  Memory Efficient Local Outlier Detection in Data Streams},'' \emph{IEEE
  Transactions on Knowledge and Data Engineering}, vol.~28, no.~12, pp.
  3246--3260, 2016.

\bibitem{gupta2014outlier}
M.~Gupta, J.~Gao, C.~Aggarwal, and J.~Han, ``{Outlier Detection for Temporal
  Data: A Survey},'' \emph{IEEE Transactions on Knowledge and Data
  Engineering}, vol.~26, no.~9, pp. 2250--2267, 2014.

\bibitem{rabin2005exchange}
M.~Rabin, ``{How To Exchange Secrets with Oblivious Transfer},'' Cryptology
  ePrint Archive, Report 2005/187, 2005.

\bibitem{angiulli2007detecting}
F.~Angiulli and F.~Fassetti, ``{Detecting Distance-Based Outliers in Streams of
  Data},'' in \emph{{CIKM}'07}, 2007.

\bibitem{kontaki2011continuous}
M.~Kontaki, A.~Gounaris, A.~Papadopoulos, K.~Tsichlas, and Y.~Manolopoulos,
  ``{Continuous Monitoring of Distance-Based Outliers over Data Streams},'' in
  \emph{{IEEE ICDE}'11}, 2011.

\bibitem{bhaduri2011privacy}
K.~Bhaduri, M.~Stefanski, and A.~Srivastava, ``{Privacy-Preserving Outlier
  Detection through Random Nonlinear Data Distortion},'' \emph{IEEE
  Transactions on Systems, Man, and Cybernetics, Part B (Cybernetics)},
  vol.~41, no.~1, pp. 260--272, 2011.

\bibitem{bohler2017privacy}
J.~B{\"o}hler, D.~Bernau, and F.~Kerschbaum, ``{Privacy-Preserving Outlier
  Detection for Data Streams},'' in \emph{{DBSec}'17}, 2017.

\bibitem{erfani2014privacy}
S.~Erfani, Y.~Law, S.~Karunasekera, A.~Leckie, and M.~Palaniswami,
  ``{Privacy-Preserving Collaborative Anomaly Detection for Participatory
  Sensing},'' in \emph{{PAKDD}'14}, 2014.

\bibitem{alabdulatif2017privacy}
A.~Alabdulatif, H.~Kumarage, I.~Khalil, and X.~Yi, ``{Privacy-Preserving
  Anomaly Detection in Cloud with Lightweight Homomorphic Encryption},''
  \emph{Journal of Computer and System Sciences}, vol.~90, pp. 28--45, 2017.

\bibitem{li2015privacy}
L.~Li, L.~Huang, W.~Yang, X.~Yao, and A.~Liu, ``{Privacy-Preserving LOF Outlier
  Detection},'' \emph{Knowledge and Information Systems}, vol.~42, no.~3, pp.
  579--597, 2015.

\bibitem{vaidya2004privacy}
J.~Vaidya and C.~Clifton, ``{Privacy-Preserving Outlier Detection},'' in
  \emph{{ICDM}'04}, 2004.

\bibitem{cao2014scalable}
L.~Cao \emph{et~al.}, ``{Scalable Distance-Based Outlier Detection over
  High-Volume Data Streams},'' in \emph{{IEEE ICDE}'14}, 2014.

\bibitem{grubbs2017leakage}
P.~Grubbs, K.~Sekniqi, V.~Bindschaedler, M.~Naveed, and T.~Ristenpart,
  ``{Leakage-Abuse Attacks against Order-Revealing Encryption},'' in
  \emph{{IEEE S\&P}'17}, 2017.

\bibitem{kornaropoulosdata}
E.~Kornaropoulos, C.~Papamanthou, and R.~Tamassia, ``{Data Recovery on
  Encrypted Databases with k-Nearest Neighbor Qquery Leakage},'' in \emph{{IEEE
  S\&P}'19}, 2019.

\bibitem{tran2016distance}
L.~Tran, L.~Fan, and C.~Shahabi, ``{Distance-Based Outlier Detection in Data
  Streams},'' \emph{Proceedings of the VLDB Endowment}, vol.~9, no.~12, pp.
  1089--1100, 2016.

\bibitem{pullonen2012design}
P.~Pullonen, D.~Bogdanov, and T.~Schneider, ``{The Design and Implementation of
  A Two-Party Protocol Suite for Sharemind 3},''
  \url{http://tubiblio.ulb.tu-darmstadt.de/61259/}[online], 2012.

\bibitem{demmler2015aby}
D.~Demmler, T.~Schneider, and M.~Zohner, ``{ABY-A Framework for Efficient
  Mixed-Protocol Secure Two-Party Computation},'' in \emph{NDSS'15}, 2015.

\bibitem{beaver1991efficient}
D.~Beaver, ``{Efficient Multiparty Protocols using Circuit Randomization},'' in
  \emph{{CRYPTO}'91}, 1991.

\bibitem{yao1982protocols}
A.~Yao, ``{Protocols for Secure Computations},'' in \emph{{IEEE SFCS}'82},
  1982.

\bibitem{bellare2012foundations}
M.~Bellare, V.~Hoang, and P.~Rogaway, ``{Foundations of Garbled Circuits},'' in
  \emph{{ACM CCS}'12}, 2012.

\bibitem{asharov2013more}
G.~Asharov, Y.~Lindell, T.~Schneider, and M.~Zohner, ``{More Efficient
  Oblivious Transfer and Extensions for Faster Secure Computation},'' in
  \emph{{ACM CCS}'13}, 2013.

\bibitem{Agent2018innoVi}
AgentVi, ``{innoVi Enterprise},''
  \url{https://www.agentvi.com/products/innovi/}\\\url{innovi-enterprise/}[online],
  2018.

\bibitem{Azure2016HoneyWell}
Microsoft, ``{Tracking a Building's Vital Signs to Keep it Safe and Healthy},''
  \url{https://customers.microsoft.com/en-us/story/tracking-a-buildings-vital-signs-to-keep-it-safe-and-h}[online],
  2016.

\bibitem{Amazon2018Kinesis}
Amazon, ``{Amazon Kinesis},'' \url{https://aws.amazon.com/kinesis/}[online],
  2018.

\bibitem{Microsoft2018Anomaly}
Microsoft, ``{Time Series Anomaly Detection},''
  \url{https://docs.microsoft.com}\\\url{/en-us/azure/machine-learning/studio-module-reference/time-series-}\\\url{anomaly-detection\#how-to-configure-time-series-anomaly-detection}
  [online], 2018.

\bibitem{kamara2011outsourcing}
S.~Kamara, P.~Mohassel, and M.~Raykova, ``{Outsourcing Multi-Party
  Computation},'' {Cryptology ePrint Archive, Report 2011/272}, 2011.

\bibitem{mohassel2017secureml}
P.~Mohassel and Y.~Zhang, ``{SecureML: A System for Scalable Privacy-Preserving
  Machine Learning},'' in \emph{{IEEE S\&P}'17}, 2017.

\bibitem{nikolaenko2013privacy}
V.~Nikolaenko \emph{et~al.}, ``{Privacy-Preserving Matrix Factorization},'' in
  \emph{{ACM CCS}'13}, 2013.

\bibitem{nikolaenko2013ridge}
------, ``{Privacy-Preserving Ridge Regression on Hundreds of Millions of
  Records},'' in \emph{{IEEE S\&P}'13}, 2013.

\bibitem{lai2019graph}
S.~Lai, X.~Yuan, S.-F. Sun, J.~K. Liu, Y.~Liu, and D.~Liu, ``{GraphSE{${}^2$}:
  An Encrypted Graph Database for Privacy-Preserving Social Search},'' in
  \emph{{ACM ASIACCS}'19}, 2019.

\bibitem{chan2005modeling}
P.~Chan and M.~Mahoney, ``{Modeling Multiple Time Series for Anomaly
  Detection},'' in \emph{{IEEE ICDE}'05}, 2005.

\bibitem{ramaswamy2000efficient}
S.~Ramaswamy, R.~Rastogi, and K.~Shim, ``{Efficient Algorithms for Mining
  Outliers from Large Data Sets},'' in \emph{{ACM SIGMOD}'00}, 2000.

\bibitem{batcher1968sorting}
K.~Batcher, ``{Sorting Networks and their Applications},'' in \emph{{ACM
  SJCC}'68}, 1968.

\bibitem{luby1988construct}
M.~Luby and C.~Rackoff, ``{How to Construct Pseudorandom Permutations from
  Pseudorandom Functions},'' \emph{SIAM Journal on Computing}, vol.~17, pp.
  373--386, 1988.

\bibitem{bost2015machine}
R.~Bost, R.~Popa, S.~Tu, and S.~Goldwasser, ``{Machine Learning Classification
  over Encrypted Data},'' in \emph{NDSS'15}, 2015.

\bibitem{canetti2000security}
R.~Canetti, ``{Security and Composition of Multiparty Cryptographic
  Protocols},'' \emph{Journal of Cryptology}, vol.~13, no.~1, pp. 143--202,
  2000.

\bibitem{xiao2018gc}
X.~Wang, ``{FlexSC},'' \url{https://github.com/wangxiao1254/FlexSC}[online],
  2018.

\bibitem{Dua:2019}
D.~Dua and C.~Graff, ``{{UCI} Machine Learning Repository},''
  \url{http://archive.ics.uci.edu/ml}[online], 2017.

\end{thebibliography}

\end{document}